\documentclass[12pt, a4paper]{article}
\pdfoutput=1
\usepackage{graphicx}
\usepackage{amssymb}
\usepackage{amsmath}
\usepackage{bm}
\usepackage{color}
\usepackage{theorem}
\usepackage{subcaption}
\usepackage{colortbl}
\usepackage{listings}
\usepackage{tabularx}
\usepackage{longtable}
\usepackage[utf8]{inputenc}

\usepackage[sort&compress,numbers, merge]{natbib}

\definecolor{Orange}{cmyk}{0,0.61,0.87,0}
\definecolor{JungleGreen}{cmyk}{0.99,0,0.52,0}
\definecolor{OliveGreen}{cmyk}{0.64,0,0.95,0.40}
\definecolor{Brown}{cmyk}{0,0.70,1,0.40}
\definecolor{RoyalBlue}{cmyk}{0.71,0.53,0,0.12}
\definecolor{Gray}{cmyk}{0,0,0,0.40}
\definecolor{LightPink}{cmyk}{0.0,0.25,0,0}
\definecolor{LLightPink}{cmyk}{0.0,0.10,0,0}
\definecolor{LightBlue}{cmyk}{0.25,0,0,0}
\definecolor{LightGray}{cmyk}{0,0,0,0.2}

\newcommand{\Slash}[1]{{\ooalign{\hfil/\hfil\crcr$#1$}}}

\newcommand{\beq}{\begin{equation}}
\newcommand{\eeq}{\end{equation}}

\allowdisplaybreaks[1]

\usepackage[colorlinks=true, linkcolor=OliveGreen, citecolor=RoyalBlue,
urlcolor=Brown]{hyperref}

\begin{document}

\begin{titlepage}

\begin{flushright}
{\tt 
KCL-PH-TH/2023-27, CERN-PH-TH-2023-083\\
FTPI-MINN-23/08,
UMN-TH-4214/23
}
\end{flushright}

\vskip 1.35cm
\begin{center}

{\large
{\bf
Electroweak Loop Contributions to the
Direct Detection of Wino Dark Matter
}
}

\vskip 1.5cm

{\bf John Ellis}$^{a}$,
{\bf Natsumi Nagata}$^{b}$, 
{\bf Keith A. Olive}$^{c}$,
and 
{\bf Jiaming Zheng}$^{d}$

\vskip 0.8cm

{\it $^a$Theoretical Particle Physics and Cosmology Group, Department of
  Physics, King's~College~London, London WC2R 2LS, United Kingdom;\\
Theoretical Physics Department, CERN, CH-1211 Geneva 23,
  Switzerland}\\[3pt]
{\it $^b$Department of Physics, University of Tokyo, Bunkyo-ku, Tokyo
 113--0033, Japan} \\[3pt]
{\it $^c$William I. Fine Theoretical Physics Institute, School of
 Physics and Astronomy, \\ University of Minnesota, Minneapolis,
 Minnesota 55455, USA} \\
 {\it $^d${Tsung-Dao Lee Institute \& School of Physics and Astronomy, \\ Shanghai Jiao Tong University, Shanghai 200240, China} }

\date{\today}

\vskip 1.5cm

\begin{abstract}
Electroweak loop corrections to the matrix elements for the spin-independent
scattering of cold dark matter particles on nuclei are generally small, typically below
the uncertainty in the local density of cold dark matter. However, as shown in this paper,
there are instances in which the electroweak loop corrections are relatively large, and
change significantly the spin-independent dark matter scattering rate. An important
example occurs when the dark matter particle is a wino, e.g., in anomaly-mediated
supersymmetry breaking (AMSB) and pure gravity mediation (PGM) models. We find that the
one-loop electroweak corrections to the spin-independent wino LSP scattering cross section generally interfere constructively with the tree-level contribution for AMSB models with negative Higgsino mixing, $\mu < 0$, and in
PGM-like models for both signs of $\mu$, lifting the cross section out of the neutrino fog
and into a range that is potentially detectable in the next generation of direct searches for
cold dark matter scattering.

\end{abstract}

\end{center}
\end{titlepage}

\section{Introduction}

As a general rule, electroweak loop corrections to the
matrix elements for the scattering of dark matter
particles on nuclei are expected to be small compared
to other uncertainties such as those in the local density
of dark matter, its velocity spectrum and the distributions
of quark and gluon constituents in nuclear matter.
However, there are instances in which electroweak loops
can make non-negligible contributions to the dark matter
scattering matrix elements, as we discuss in this paper,
focusing on the case of the scalar matrix elements that 
dominate spin-independent dark matter scattering.

These instances arise when the tree-level scattering
matrix element is suppressed, e.g., because scalar
exchange is parametrically reduced as in the case
of a wino-like dark matter particle that couples
weakly to an intermediate Higgs boson, or because of an
accidental cancellation for specific values of the
contributions to the hadronic scattering matrix element
that causes a `blind spot' in the model parameter space.
The former case is relevant, in particular, in models
with anomaly-mediated supersymmetry breaking (AMSB) \cite{anom,mAMSB,mc-amsb,ehow++} or
pure gravity mediation (PGM) \cite{pgm,eioy,PGMh}, which predict that the
lightest supersymmetric particle (LSP) is a wino-like
neutralino. A first example of a `blind spot' was given
in the MSSM with supersymmetry-breaking parameters
constrained to be universal at the input grand unification
scale (the CMSSM) for a negative sign of the Higgsino
mixing term $\mu$ \cite{Falk:1998xj,cancel,Cheung:2012qy}.

A lot of experimental water has passed under the
supersymmetric bridge since the CMSSM, AMSB and PGM
were initially formulated, with the LHC and direct
searches for dark matter via scattering experiments
taking their tolls on the respective supersymmetric
model parameter spaces \cite{nosusy}.
For example, a recent survey of the CMSSM \cite{Ellis:2022emx} found allowed
strips of parameter space in which the dark matter density
was brought into the allowed range by either the focus-point
mechanism \cite{fp} or stop coannihilation \cite{stopco},
for Higgsino or bino LSP with Higgs mass calculations
enforcing a heavy spectrum.
A global analysis of the AMSB model that took into 
account LHC and other experimental constraints \cite{mc-amsb} found
them to be consistent with a wino LSP weighing 
$\sim 3$~TeV \cite{winomass1,winomass} that had a very low spin-independent
dark matter scattering cross section that could descend
into the neutrino background `fog' \cite{floor} \footnote{We note that several studies have argued that wino dark matter may be in conflict with Fermi-LAT and H.E.S.S. observations of the gamma-rays coming from the Galactic centre and dwarf spheroidal galaxies \cite{gamma}. The degree of this tension depends, however, on the uncertainties in the dark matter density profiles in these objects.}. However, this
analysis did not take into account electroweak loop
corrections to the spin-independent scattering matrix element,
which can become important in regions where the tree-level
scattering matrix element is suppressed, e.g., because of large
sparticle masses and/or a chance cancellation.

We recall that in mAMSB models \cite{mAMSB} the number of free parameters is reduced relative to the CMSSM,
from four to three. The gravitino mass, $m_{3/2}$ provides a seed for gaugino masses and $A$-terms, which are generated by radiative corrections. For example, at one loop, the gaugino masses at some high-energy scale (often taken to be the GUT scale as in the CMSSM) are given by~\cite{anom}:
\begin{eqnarray}
    M_{1} &=&
    \frac{33}{5} \frac{g_{1}^{2}}{16 \pi^{2}}
    m_{3/2}\ ,
    \label{eq:M1} \\
    M_{2} &=&
    \frac{g_{2}^{2}}{16 \pi^{2}} m_{3/2}  \ ,
        \label{eq:M2}     \\
    M_{3} &=&  -3 \frac{g_3^2}{16\pi^2} m_{3/2}\ ,
    \label{eq:M3}
\end{eqnarray}
with $M_i \ll m_{3/2}$. This results in a mass spectrum with $|M_1| : |M_2| : |M_3| \approx 2.8:1:7.1$, resulting in a wino LSP if the scalar masses are sufficiently heavy. Similar one-loop expressions determine the trilinear terms at the same high-energy input scale. Requiring that scalar masses are also determined radiatively results in an unrealistic model, because renormalization leads to negative squared masses for sleptons. Thus the minimal AMSB scenario (mAMSB) adds a constant $m_0^2$ to all  squared scalar masses~\cite{mAMSB}. Thus the mAMSB model has three continuous free parameters: $m_{3/2}$, $m_0$ and the ratio of Higgs vevs, $\tan \beta$. In the limit that $m_0 = m_{3/2}$, one recovers the set of two-parameter PGM models \cite{eioy}, specified by $m_{3/2}$ and $\tan \beta$. For these models to be viable, $\tan \beta \approx 2$ is required. As in the CMSSM, the $\mu$-term and the Higgs pseudoscalar mass (or $B_0 \mu$) are determined from the minimization of the Higgs potential, and the sign of $\mu$ is free. 

In both the mAMSB~\cite{mAMSB} and PGM-like~\cite{PGMh} models the LSP is typically a
wino-like neutralino \footnote{Although there are limiting cases in both models
where the LSP becomes Higgsino-like.}. In such scenarios, if the Higgsino
mixing parameter $\mu \gg M_2$ and $m_Z$, the tree-level 
Higgs exchange contribution to spin-independent dark matter
scattering is strongly suppressed, which is why the 
electroweak loop corrections can be important. This is the case, in particular, when
there is a cancellation in the tree-level scattering matrix element, as can occur in
both the mAMSB and PGM-like models, as we discuss here.

The outline of this paper is as follows. In Section~\ref{sec:spin-independentcalx} we
discuss calculations of the spin-independent dark matter scattering cross section,
first reviewing the relevant effective interactions and then the tree-level contribution of 
Higgs exchange and then the form of the one-loop electroweak radiative corrections.
In Section~\ref{sec:relicmh} we discuss calculations of two auxiliary quantities, namely
the relic wino dark matter density and the Higgs mass, paying particular attention to the 
requirements of radiative electroweak symmetry breaking and a reliable calculation of the
Higgs mass in supersymmetric models with a very heavy spectrum. We present our results
for the spin-independent wino dark matter scattering cross section in Section~\ref{sec:results},
exhibiting the regions of AMSB and PGM-like parameter space where the one-loop effects
enhance the cross section, as well as regions where they suppress it. For reasons
that we explain, both effects are possible in the AMSB model, whereas the cross section is
generally enhanced in the PGM-like model. As we discuss in Section~\ref{sec:conx}, 
there are generic regions of the
models' parameter spaces where the enhanced cross section rises out of the neutrino fog and 
may be detectable in the next generation of direct searches for cold dark matter scattering.

\section{Spin-Independent Scattering Cross Sections}
\label{sec:spin-independentcalx}

\subsection{Effective interactions}

We first review the calculation of the cross section for the elastic scattering 
of wino-like dark matter on a nucleus. We focus on the spin-independent
elastic scattering, as the spin-dependent scattering cross section turns
out to be negligibly small for the cases considered here. The spin-independent scattering cross
section for generic Majorana fermion dark matter is given by
\cite{GW,kg,bfg,Ellis:1992ka,dn,GJK,Falk:1998xj} 
\begin{equation}
 \sigma_{\text{SI}}^{Z, A} = \frac{4}{\pi} \biggl(
\frac{m_\chi m_T}{m_\chi+ m_T}
\biggr)^2 
\left[Z f_p + (A-Z) f_n\right]^2 ~,
\end{equation}
where $m_\chi$ and $m_T$ are the masses of the Majorana dark matter and
the target nucleus, respectively, $A$ and $Z$ are the mass and atomic
numbers of the nucleus, and $f_N$ $(N= p,n)$ are the effective dark matter-nucleon
couplings.

The effective couplings $f_N$ are obtained as a sum of the Wilson
coefficients of dark matter-quark/gluon effective operators multiplied by
their nucleon matrix elements. The operators relevant to our discussions
are \cite{dn, Falk:1998xj, efso, Hisano:2010fy, Hisano:2010ct, Hisano:2015bma}
\begin{align}
 {\cal L}_{\text{eff}}
&= \sum_{q} \alpha_{3q}\, \overline{\tilde{\chi}^0} \tilde{\chi}^0 \bar{q}q +
\alpha_G\, \frac{\alpha_s}{\pi} \overline{\tilde{\chi}^0} \tilde{\chi}^0
 G^a_{\mu\nu} G^{a\mu\nu}
\nonumber \\
&+ \sum_{q } \frac{\beta_{1q}}{m_\chi} \, \overline{\tilde{\chi}^0} i \partial^\mu
 \gamma^\nu\tilde{\chi}^0 \, {\cal O}^q_{\mu\nu}   
+ \sum_{q } \frac{\beta_{2q}}{m_\chi^2} \, \overline{\tilde{\chi}^0} i \partial^\mu
 i \partial^\nu \tilde{\chi}^0 \, {\cal O}^q_{\mu\nu} ~,
\label{eq:lageff}
\end{align}
where $\tilde{\chi}^0$ denotes the dark matter field, the $q$ are quark fields,
$G^a_{\mu\nu}$ is the field strength tensor of the gluon, $\alpha_s \equiv
g_s^2/(4\pi)$ is the strong gauge coupling constant, and the ${\cal
O}^q_{\mu\nu}$ are the so-called twist-2 operators of the quark
fields~\footnote{One could also consider the interaction described by the
gluonic twist-2 operator. However, as we see below, this interaction
is induced at higher order in $\alpha_s$ compared with the other
interactions, and thus can be neglected for the leading order computation.}
defined by \cite{dn, GJK}
\begin{equation}
 {\cal O}^q_{\mu\nu} \equiv \frac{1}{2} \bar{q} i \Bigl(
D_\mu \gamma_\nu + D_\nu \gamma_\mu - \frac{1}{2}\eta_{\mu\nu} \Slash{D}
\Bigr) q ~,
\end{equation}
with $D_\mu$ denoting the covariant derivative.

\subsection{Tree-level Higgs exchange}

\begin{figure}
\centering
\includegraphics[height=45mm]{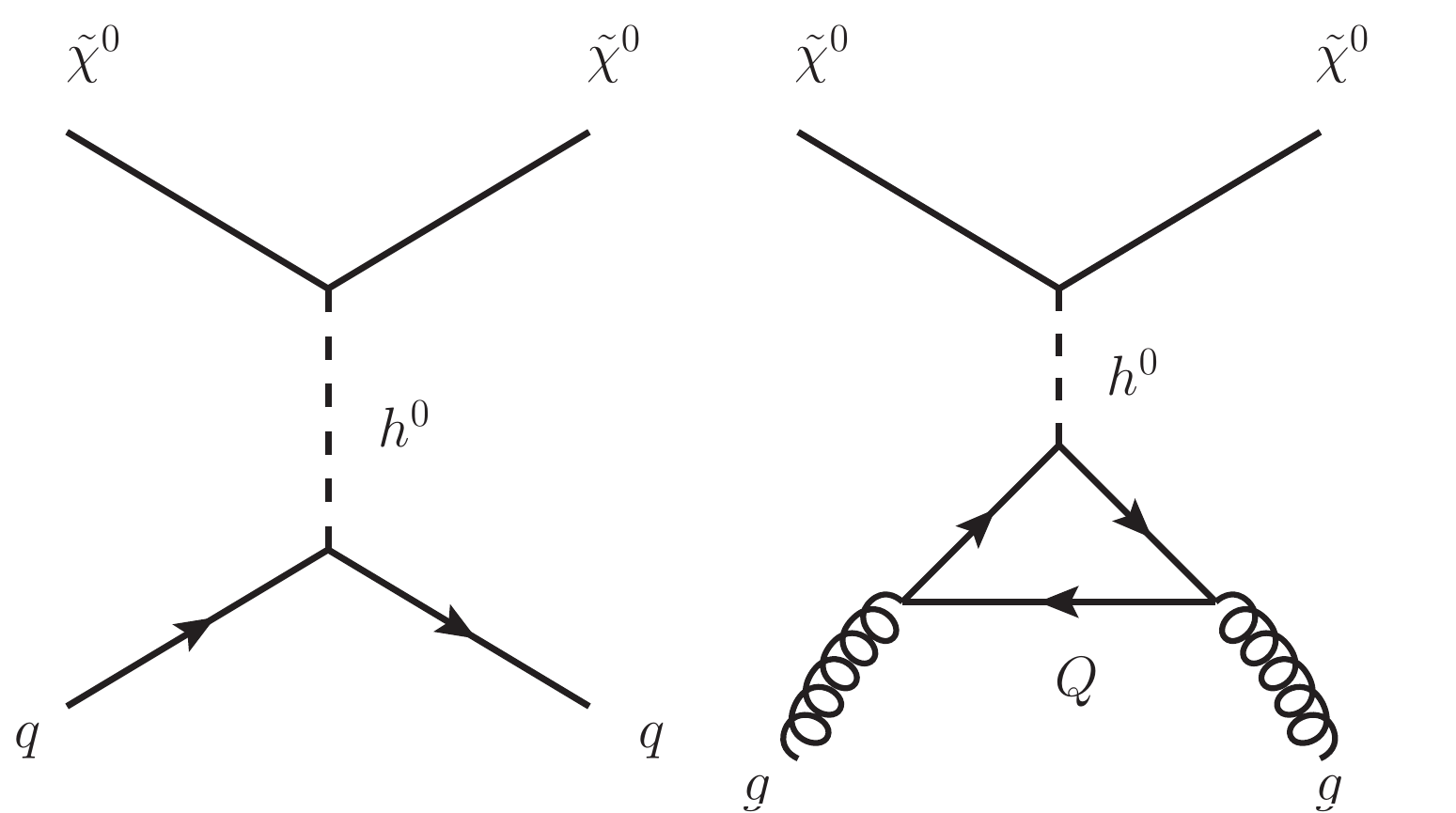}
\caption{\it 
The Higgs-boson exchange processes that contribute at leading order to the
spin-independent scattering of wino-like neutralino with nucleons.
}  
  \label{fig:tree}
\end{figure}
The effective four-fermion coupling, $\alpha_{3q}$ in Eq.~(\ref{eq:lageff}) contains contributions from both squark and Higgs exchange. However, in the region of parameter space of interest here, the squark masses and the heavy Higgs scalar mass, $m_H$, are significantly larger than the light Higgs scalar mass, $m_h$. Thus, 
at leading order, wino-like neutralino dark matter-nucleon scattering
is induced by the light Higgs-boson exchange processes shown in
Fig.~\ref{fig:tree}. By integrating out the Higgs boson, we can readily
obtain the effective quark couplings $\alpha_{3q}$ \cite{Falk:1998xj,
efso} 
\begin{equation}
 \alpha_{3q}^{(\text{tree})} \simeq \frac{gm_q}{4 m_W m_h^2}
\text{Re} \left[
\left(Z_{\chi 3} \cos \beta - Z_{\chi 4} \sin \beta \right)
\left(g Z_{\chi 2} - g^\prime Z_{\chi 1}\right)
\right] ~,
\label{eq:a3aprx1}
\end{equation}
where $g$ and $g^\prime$ denote the SU(2)$_L$ and U(1)$_Y$ gauge
couplings, respectively, $m_W$ is the $W$-boson mass, the $m_q$ are the
quark masses, and $\tan \beta$ is the ratio of the Higgs VEVs defined by
$\tan\beta \equiv \langle H_u^0 \rangle/\langle H_d^0 \rangle$. Note that in the assumed limit, $m_H \gg m_h \simeq 125$ GeV, the Higgs mixing angle, $\alpha$ can be approximated by the decoupling limit $\alpha \to \beta - \pi/2$,
so that $\cos \alpha \to \sin \beta$ and $\sin \alpha \to - \cos
\beta$. The $Z_{\chi i}$ ($i = 1,2,3,4$) in Eq.~\eqref{eq:a3aprx1} are
defined by
\begin{equation}
 \tilde{\chi}^0 = Z_{\chi 1} \tilde{B} + Z_{\chi 2} \tilde{W}^0 +
Z_{\chi 3} \tilde{H}_d^0 + Z_{\chi 4} \tilde{H}^0_u ~, 
\end{equation}
where $\tilde{B}$, $\tilde{W}^0$, $\tilde{H}^0_d$, and $\tilde{H}^0_u$,
denote the bino, wino, and Higgsino fields, respectively. Focusing on the 
region of parameter space in which $\tilde{\chi}^0$ is
wino-like and $m_W \ll |M_i - \mu|$ ($i = 1,2$), where $M_1$, $M_2$ and
$\mu$ are the bino mass, wino mass and Higgsino mixing parameters, respectively, we 
have
\begin{eqnarray}
 Z_{\chi 1} \simeq 0, \quad
 Z_{\chi 2} \simeq 1, & &
 Z_{\chi 3} \simeq \frac{m_W}{M_2^2 - \mu^2} (M_2 \cos \beta + \mu
 \sin\beta), \nonumber \\
 Z_{\chi 4} & \simeq & -\frac{m_W}{M_2^2 - \mu^2} (M_2 \sin \beta + \mu
 \cos\beta) \, ,
\label{eq:zchiaprx}
\end{eqnarray}
and we can further approximate Eq.~\eqref{eq:a3aprx1} as
\begin{equation}
 \alpha_{3q}^{(\text{tree})} \simeq
 \frac{g^2 m_q (M_2 + \mu \sin 2\beta)}{4 m_h^2 (M_2^2 -\mu^2)} ~.
\label{eq:a3aprx2}
\end{equation}
From this expression, we see that $\alpha_{3q}^{(\text{tree})}$
decreases as $\propto \mu^{-1}$ when $|\mu| \gg M_2$, which is the case
for the AMSB so long as $m_0 \ll m_{3/2}$. These approximations also hold in the PGM model for values of $\tan \beta$ that maintain $|\mu| \gg M_2$.

The contribution of the diagrams in Fig.~\ref{fig:tree} to the effective
nucleon couplings is obtained by using the nucleon matrix elements of
the quark-antiquark scalar operators, $\bar{q}q$. Their values are often described
by the mass fractions $f_{T_q}^{(N)}$ defined by 
\beq
m_N f_{T_q}^{(N)} \equiv
\langle N | m_q \bar{q}q |N\rangle \equiv \sigma_q \equiv m_q B_q^N \, , 
\label{massfractions}
\eeq
with $m_N$ the nucleon mass. 
For the light quarks, combinations of the matrix elements can be related to $\sigma$ terms such as
the
pion-nucleon $\sigma$ term
\begin{equation}
\Sigma_{\pi N} \; = \; \frac{1}{2} (m_u + m_d) \left( B_u^p + B_d^p \right) \, ,
\label{piNSigma}
\end{equation}
which may be computed directly with lattice simulations~\cite{FlavourLatticeAveragingGroupFLAG:2021npn} or  extracted phenomenologically from data on low-energy $\pi$-nucleon scattering
or on pionic atoms~\cite{atoms}. 
Another combination  can be extracted phenomenologically from the octet baryon mass splittings,
\begin{equation}
\sigma_0 \; = \; \frac{1}{2} (m_u + m_d) \left( B_u^p + B_d^p - 2 B_s^p\right) \, ,
\label{sigma0}
\end{equation}
which gives
\beq
\sigma_s  = m_s B^p_s = \frac{m_s}{m_u+m_d} (\Sigma_{\pi N} -\sigma_0)\, .
\label{sigmas}
\eeq
A third relation is obtained from baryon octet masses 
\begin{equation}
z \; \equiv \; \frac{B_u^p - B_s^p}{B_d^p - B_s^p} \; = \; \frac{m_{\Xi^0} + m_{\Xi^-} - m_p - m_n}{m_{\Sigma^+} + m_{\Sigma^-} - m_p - m_n} \; = \; 1.49 \, .
\label{zratio}
\end{equation}
From a recent compilation in
Ref.~\cite{Ellis:2018dmb}, we use 
\begin{equation}
 \Sigma_{\pi N} \; = \; 46 \pm 11 \; {\rm MeV} \qquad \sigma_s \;  = \;  35 \pm 16 \; {\rm MeV}  \, ,
\label{95percent}
\end{equation}
which determines the matrix elements of the three light quarks and we
list them in the first three columns of Table~\ref{tab:fTvalues} for the
reader's convenience. 

\begin{table}[!ht]
\caption{\it Values of the mass fractions $f^{(N)}_{T_{q,G}}$ obtained in
 Ref.~\cite{Ellis:2018dmb}.}
\label{tab:fTvalues}
\centering
\begin{tabular}{l||cccc|ccc}
\hline
\hline
 & $f^{(N)}_{T_u}$ & $f^{(N)}_{T_d}$ & $f^{(N)}_{T_s}$ & $f^{(N)}_{T_G}$
		 & $f^{(N)}_{T_c}$ & $f^{(N)}_{T_b}$ & $f^{(N)}_{T_t}$\\ 
\hline
Proton & 0.018(5) & 0.027(7) & 0.037(17) & 0.917(19) &0.078(2) & 0.072(2) & 0.069(1) \\
Neutron & 0.013(3) & 0.040(10) & 0.037(17) & 0.910(20) & 0.078(2) & 0.071(2) & 0.068(2) \\
\hline
\hline
\end{tabular}
\end{table}

The contributions of the heavy quarks to the scattering cross section are often calculated by integrating them out and replacing them by the one-loop gluon contributions so that $f_{T_Q}^{(N)} = \frac{2}{27} f_{T_G}^{(N)}$ for $Q = c,b,t$, where the nucleon matrix element of the gluon scalar operator
is evaluated using the trace anomaly formula given in~\cite{SVZ}:
\begin{equation}
 \langle N | \frac{\alpha_s}{\pi} G^a_{\mu\nu} G^{a\mu\nu} |N\rangle 
= - \frac{8}{9} m_N f_{T_G}^{(N)} ~,
\label{eq:gluscamat}
\end{equation}
which holds at the leading order in $\alpha_s$,
and we note in addition that $f_{T_G}^{(N)} \equiv 1-\sum_{q=u,d,s}
f_{T_q}^{(N)}$.

\if
We then find in the three quark
flavor case \textcolor{red}{???}
\begin{align}
 f_N^{(\text{tree})} &= \sum_{q = u,d,s} \frac{\alpha_{3q}^{(\text{tree})}}{m_q} 
\langle N | m_q \bar{q}q |N\rangle - \frac{1}{12} \sum_{Q=c,b,t} 
\frac{\alpha_{3Q}^{(\text{tree})}}{m_Q} 
 \langle N | \frac{\alpha_s}{\pi} G^a_{\mu\nu} G^{a\mu\nu} |N\rangle
 \nonumber \\
&= m_N 
\biggl[
\sum_{q = u,d,s} \frac{\alpha_{3q}^{(\text{tree})}}{m_q} f_{T_q}^{(N)}
+ \frac{2}{27}f_{T_G}^{(N)}  \sum_{Q=c,b,t} 
\frac{\alpha_{3Q}^{(\text{tree})}}{m_Q} 
\biggr]~,
\end{align}
where the three light quarks contribute to the effective coupling via
the tree-level process shown in the left panel in Fig.~\ref{fig:tree},
while the other three heavy quarks contribute via the one-loop diagram
shown in the right panel.
\fi

However, momenta
around the mass scale of the quark running in the loop make the most important contributions to
the integral in the loop diagram shown in Fig.~\ref{fig:tree}. These contributions are often referred to as
long-distance contributions
as the relevant energy scales are much lower than
the electroweak scale in the cases of the charm and bottom quarks. 
Since $\alpha_s (m_Q)$ is rather large at the scales of
these masses, higher-order QCD corrections are significant. We take these corrections into account to
${\cal O}(\alpha_s^3)$ in perturbative QCD, following
Ref.~\cite{vecchi}, finding
\begin{eqnarray}
  \sigma_c & = & \frac{2}{27} \left(-0.3 + 1.48 f^{(N)}_{T_G} \right) M_N  \; = \; 73.4 \pm 1.9 \; {\rm MeV} \, , \\
  \sigma_b & = & \frac{2}{27} \left(-0.16 + 1.23 f^{(N)}_{T_G} \right) M_N \; = \; 67.3 \pm 1.6 \; {\rm MeV} \, , \\
\sigma_t & = & \frac{2}{27} \left(-0.05 + 1.07 f^{(N)}_{T_G} \right) M_N \; = \; 64.7 \pm 1.4 \; {\rm MeV} 
  \, .  
\end{eqnarray}
Using (\ref{massfractions}), the corrected contributions for all quarks can then be
expressed in terms of the effective mass fractions $f^{(N)}_{T_Q}$, whose
values we show in the last three columns of Table~\ref{tab:fTvalues}. The resulting
effective coupling is given by
\begin{equation}
 f_N^{(\text{tree})} = m_N\sum_{q = u,d,s, c, b, t}
  \frac{\alpha_{3q}^{(\text{tree})}}{m_q} f_{T_q}^{(N)} ~.
\label{eq:fntree}
\end{equation}
This treatment increases the resulting value of $f_N$ by ${\cal
O}(10)$\% \cite{Ellis:2018dmb} relative to computing $f_N$ using a common $f_{T_Q}^{(N)}$ for the heavy quarks. 
We note that even though the contributions from heavy quarks are
induced by QCD loop diagrams, the resultant contributions are similar in magnitude to those from the light quarks~\cite{dn, Hisano:2010fy, Hisano:2010ct}. This is
because of the large contribution of the gluons to the mass of
nucleon: $f_{T_G}^{(N)} \gg f_{T_q}^{(N)}$.

\subsection{Electroweak loop contributions}

\begin{figure}[!ht]
\centering
\includegraphics[height=45mm]{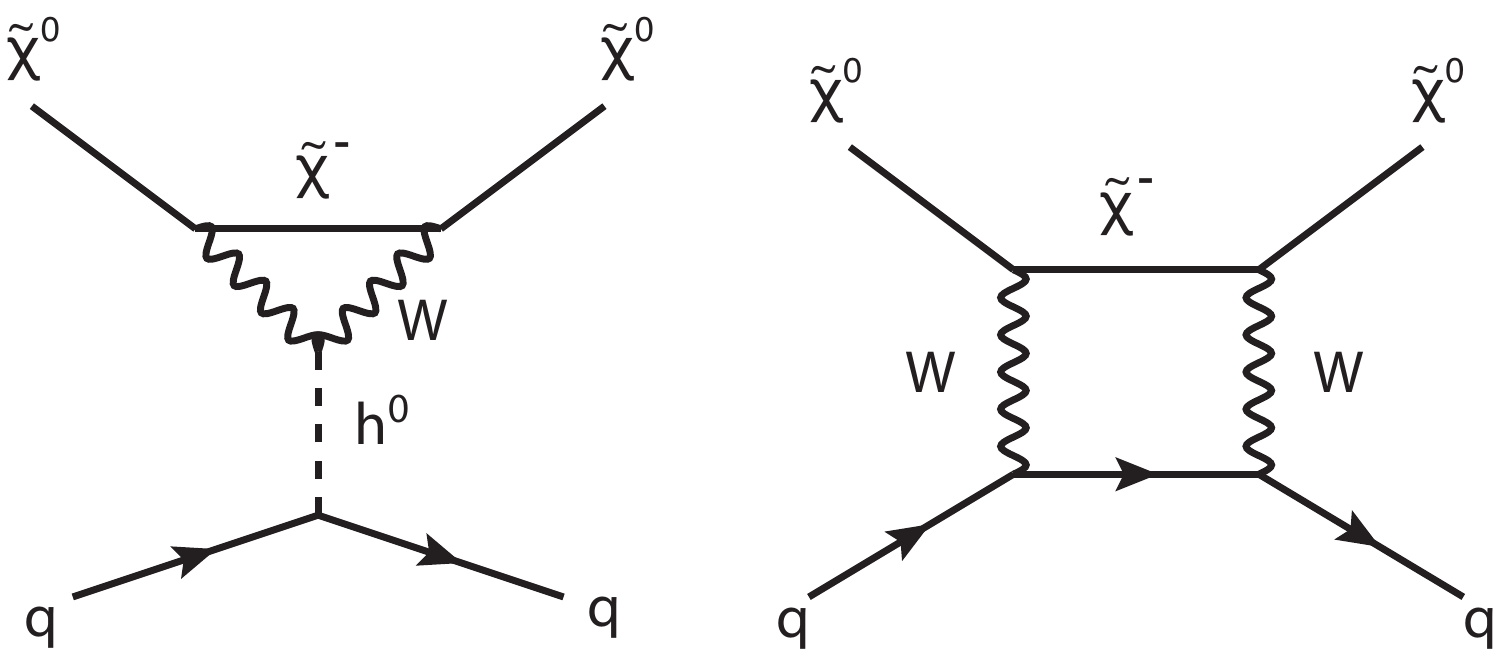}
\caption{\it 
One-loop contributions to wino-nucleon scattering.
}  
  \label{fig:wino1loop}
\end{figure}

\begin{figure}[!ht]
\centering
\includegraphics[height=55mm]{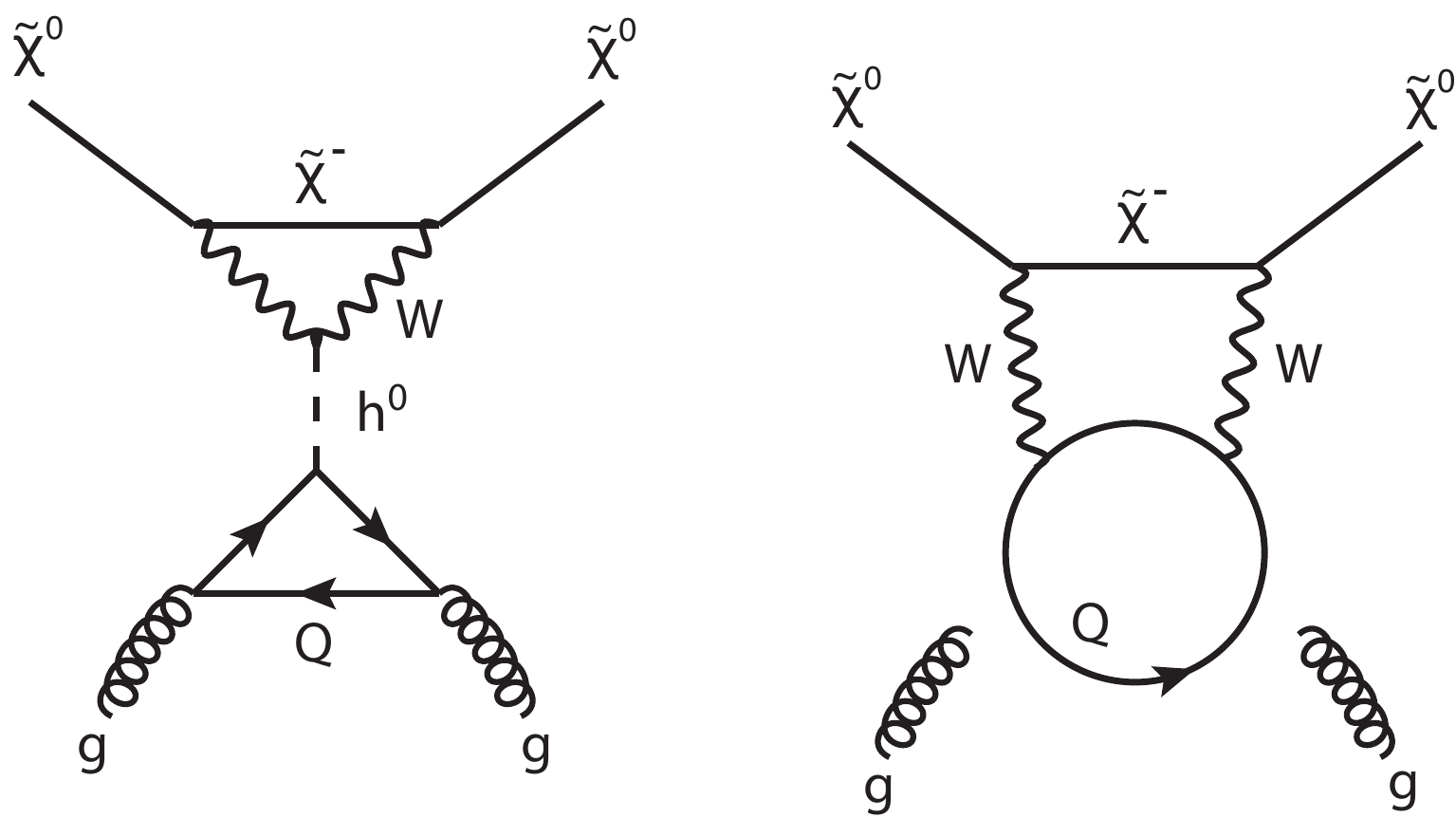}
\caption{\it 
Two-loop contributions to wino-nucleon scattering.
}  
  \label{fig:wino2loop}
\end{figure}

As seen in Eq.~\eqref{eq:a3aprx2}, the tree-level Higgs exchange
contribution is suppressed when the Higgsino mass is very large. In this
case, the contributions from the loop processes shown in
Figs.~\ref{fig:wino1loop} and \ref{fig:wino2loop} may dominate over the
tree-level contribution \cite{Hisano:2012wm}. These loop contributions have been
computed in the literature \cite{Hisano:2010fy, Hisano:2010ct,
Hisano:2011cs, Hill:2011be, Hisano:2015rsa}, and
we include them in our analysis with the following approximations:
\begin{itemize}
 \item We use the results obtained for a pure wino, though the wino-like
       neutralino LSP in our case also contains admixtures of the bino and Higgsinos. This approximation is valid since the loop
       corrections can be significant only when the Higgsino mass is
       quite large so that the tree-level bino contribution becomes
       small---in this case, $|Z_{\chi 1}|, |Z_{\chi 3}|, |Z_{\chi 4}|
       \ll 1$ as can be seen from Eq.~\eqref{eq:zchiaprx}, {i.e.},
       the LSP is almost pure wino.

 \item Although the electroweak loop contributions have been computed at
       NLO in QCD \cite{Hisano:2015rsa}, we use the LO result, since the
       difference between the two approximations can be neglected
       for our purposes.
\end{itemize}
In the rest of this subsection, we summarize our results for the
electroweak loop contributions.

The left diagram in Fig.~\ref{fig:wino1loop} gives rise to scalar 
light-quark operators, with coefficients given by 
\begin{equation}
  \alpha_{3q}^{(\text{loop})} =\frac{\alpha_2^2 m_q}{4m_W m_{h}^2} g_{\rm
   H}(\omega) ~,
\label{eq:alp3qloop}
\end{equation}
where $\alpha_2\equiv g^2/(4\pi)$ is the SU(2)$_L$ coupling strength and
$\omega \equiv m_W^2/m_\chi^2$. The mass function  $g_{\text{H}} (x)$ is
given in the Appendix. In addition, heavy quarks provide scalar
gluon operator contributions via the two-loop diagrams in
Fig.~\ref{fig:wino2loop}. As in the case of the tree-level Higgs
exchange processes, these two-loop contributions can be comparable to the
one-loop contribution to the scalar quark operators, even though they are
induced at a higher order in ${\cal O}(\alpha_s)$. 

We again take account
of the long-distance QCD corrections to the left diagram in
Fig.~\ref{fig:wino2loop} by using the $f^{(N)}_{T_Q}$ $(Q = c,b,t)$ values given
in Table~\ref{tab:fTvalues}. For each heavy quark we have
\begin{equation}
  \alpha_{3Q}^{(\text{loop})} =\frac{\alpha_2^2 m_Q}{4m_W m_{h}^2} g_{\rm
   H}(\omega) ~,
\end{equation}
which is the same as in Eq.~\eqref{eq:alp3qloop}. 

The contribution from the right diagram in Fig.~\ref{fig:wino2loop} is
included in the coefficient $\alpha_G$ of the scalar gluon operator:
\begin{equation}
 \alpha_G =2\times 
 \frac{\alpha_2^2}{4m_W^3} g_{\rm B1}(\omega) 
+
\frac{\alpha_2^2}{4m_W^3} g_{\rm B3}(\omega,\tau)
\ ,
\label{eq:alpg}
\end{equation}
where $\tau\equiv m_t^2/m_\chi^2$ and the mass functions $g_{\text{B1}}$
and $g_{\text{B3}}$ are given in the Appendix. The
first (second) term in the right-hand side of the above expression
corresponds to the first- and second-generation
(third-generation) contribution. We use the LO formula
\eqref{eq:gluscamat} to evaluate the nucleon matrix element of the gluon
scalar operator. A more systematic treatment for the inclusion of the
higher-order QCD effects as well as the separation between the long- and
short-distance contributions using the matching procedure is discussed
in Ref.~\cite{Hisano:2015rsa}.

The right diagram in Fig.~\ref{fig:wino1loop} induces
interactions described by the twist-2 operators in
Eq.~\eqref{eq:lageff}. The Wilson coefficients of these operators are
found to be
\begin{align}
\beta_{1q}&=\frac{\alpha_2^2}{m_W^3} g_{\rm T1}(\omega) \ ,\\
\beta_{2q}&=\frac{\alpha_2^2}{m_W^3} g_{\rm T2}(\omega) \ ,
\label{eq:twist2}
\end{align}
for $q=u,d,s,c$ and
\begin{align}
\beta_{1b}&=\frac{\alpha_2^2}{m_W^3} h_{\rm T1}(\omega,\tau) \ ,\\
\beta_{2b}&=\frac{\alpha_2^2}{m_W^3} h_{\rm T2}(\omega, \tau) \ ,
\label{eq:twist2b}
\end{align}
for the $b$ quark. The nucleon matrix elements of the twist-2 operators are given
by the second moments of the parton distribution functions (PDFs) \cite{dn, GJK}:
\begin{align}
\langle N(p)\vert 
{\cal O}_{\mu\nu}^q
\vert N(p) \rangle 
&=m_N \left(
\frac{p_{\mu}p_{\nu}}{m_N^2}-\frac{1}{4}\eta_{\mu\nu}
\right)
\left[
q_N(2)+\bar{q}_N(2) 
\right]
 ~,
\label{eq:q2}
\end{align}
with
\begin{align}
q_N(2) &= \int^1_0 dx~ x\ q_N(x)~,
\\
\bar{q}_N(2) &= \int^1_0 dx~ x\ \bar{q}_N(x)~,
\end{align}
where $q_N(x)$ and $\bar{q}_N(x)$ are
the PDFs of the quark and antiquark, respectively. We present in
Table~\ref{table:2ndmoments} the values of the second moments at the
scale $\mu =m_Z$ for the proton, where we have used the CJ12
NLO PDFs given by the CTEQ-Jefferson Lab collaboration
\cite{Owens:2012bv}. As mentioned above, there is also a gluon
twist-2 contribution, but this can be neglected, as it is higher order
in $\alpha_s/\pi$, and the nucleon matrix element of the gluon twist-2 operator,
$g(2)$, is not so much larger than the light-quark operators, $u(2)$ and
$d(2)$, in contrast to the cases for the scalar operators. Thus the gluon
twist-2 contribution is always suppressed by an extra $\alpha_s/\pi$
factor compared to the light-quark twist-2 contributions.

\begin{table}
\caption{\it Second moments of the PDFs of partons in the proton evaluated at $\mu
 =m_Z$. We use the CJ12 next-to-leading order PDFs given by the
 CTEQ-Jefferson Lab collaboration \cite{Owens:2012bv}. }
\label{table:2ndmoments}
 \begin{center}
\begin{tabular}{ll|ll}
\hline
\hline
$g(2)$ & 0.464(2) & &\\
$u(2)$ & 0.223(3) &$\bar{u}(2)$ & 0.036(2) \\
$d(2)$ & 0.118(3) &$\bar{d}(2)$ & 0.037(3) \\
$s(2)$ & 0.0258(4) &$\bar{s}(2)$ & 0.0258(4) \\
$c(2)$ & 0.0187(2) &$\bar{c}(2)$ & 0.0187(2) \\
$b(2)$ & 0.0117(1) &$\bar{b}(2)$ & 0.0117(1) \\
\hline
\hline
\end{tabular}
\end{center}
\end{table}

\subsection{Summary}

Combining the results above, the effective coupling $f_N$ is evaluated as
\begin{align}
 \frac{f_N}{m_N}&= \frac{f_N^{(\text{tree})}}{m_N} +\sum_{q=u,d,s, c, b,t} 
 f_{T_q}^{(N)} \frac{\alpha_{3q}^{(\text{loop})}}{m_q}
-\frac{8}{9}f_{T_G}^{(N)} \alpha_G 
 +\sum_{q=u,d,s,c,b}\frac{3}{4} (q_N(2)+\bar{q}_N(2))(\beta_{1q} + \beta_{2q})
~ , \label{f}
\end{align}
where $f_N^{(\text{tree})}$ is given in Eq.~\eqref{eq:fntree}, the
$f_{T_q}^{(N)}$ are given in Table~\ref{tab:fTvalues},
$\alpha_{3q}^{(\text{loop})}$ is given in Eq.~\eqref{eq:alp3qloop}, $\alpha_G$
is given in Eq.~\eqref{eq:alpg}, $\beta_{1q}$ and $\beta_{2q}$ are shown
in Eqs.~\eqref{eq:twist2} and \eqref{eq:twist2b}, respectively, the second moments
$q_N(2)$ and $\bar{q}_N(2)$ are given in
Table~\ref{table:2ndmoments}, and the mass functions in the coefficients are
summarized in the Appendix.

\section{Calculation of the Relic Density and Higgs Mass}
\label{sec:relicmh}

\subsection{Relic Wino LSP Density}

\begin{figure}[!ht]
  \centering
\includegraphics[width=0.48\columnwidth]{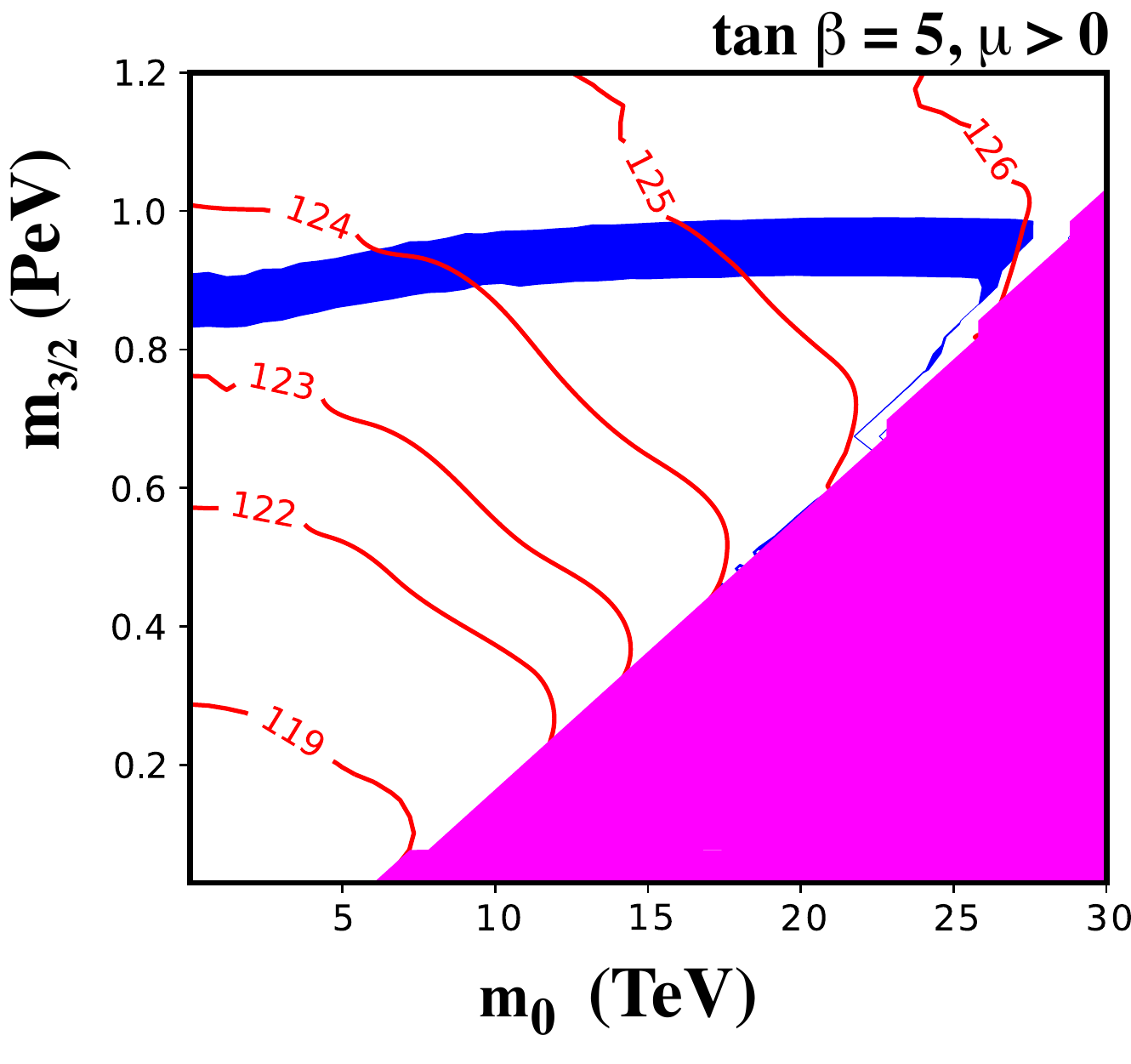}
\includegraphics[width=0.48\columnwidth]{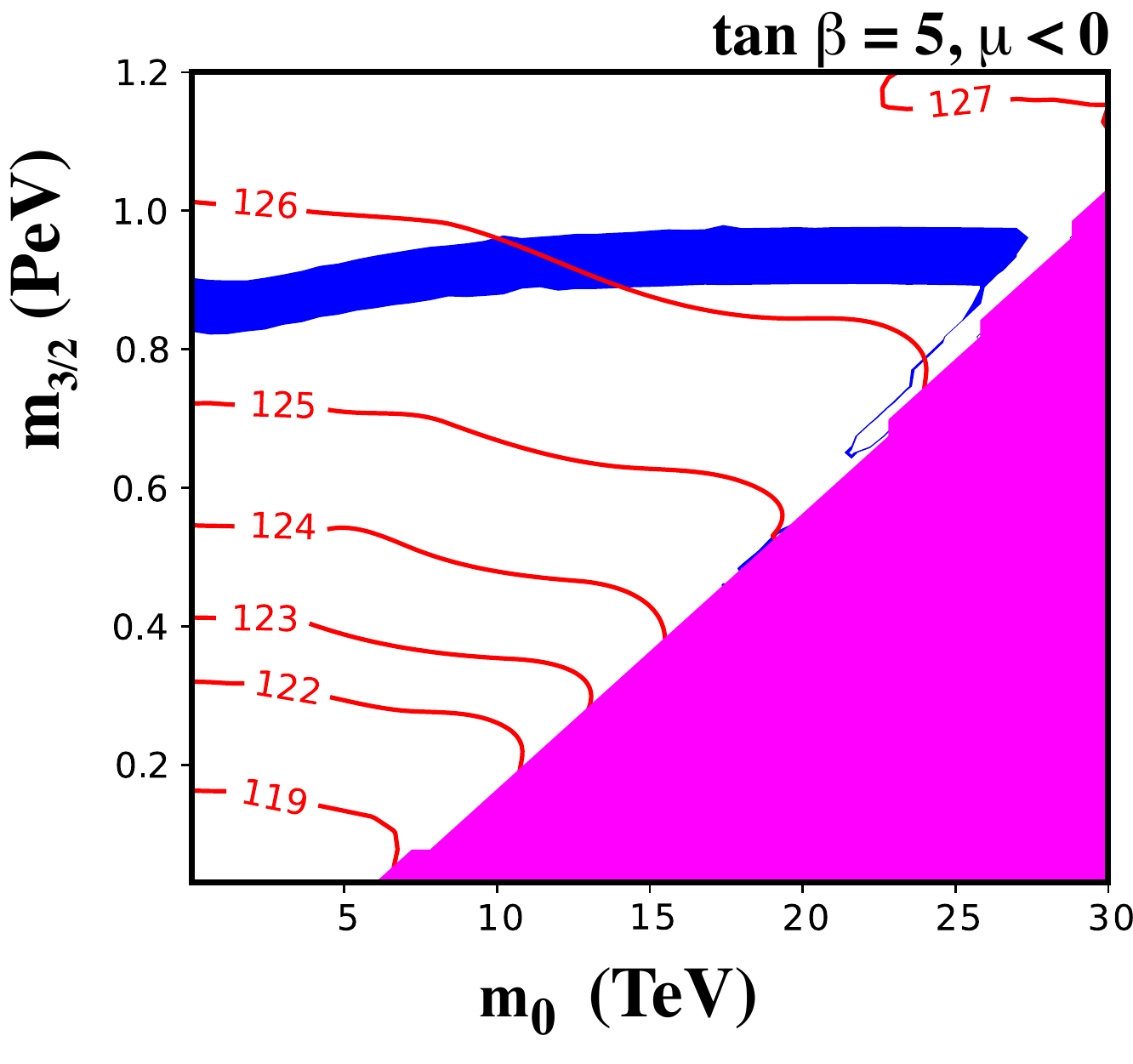} 
\\
\includegraphics[width=0.48\columnwidth]{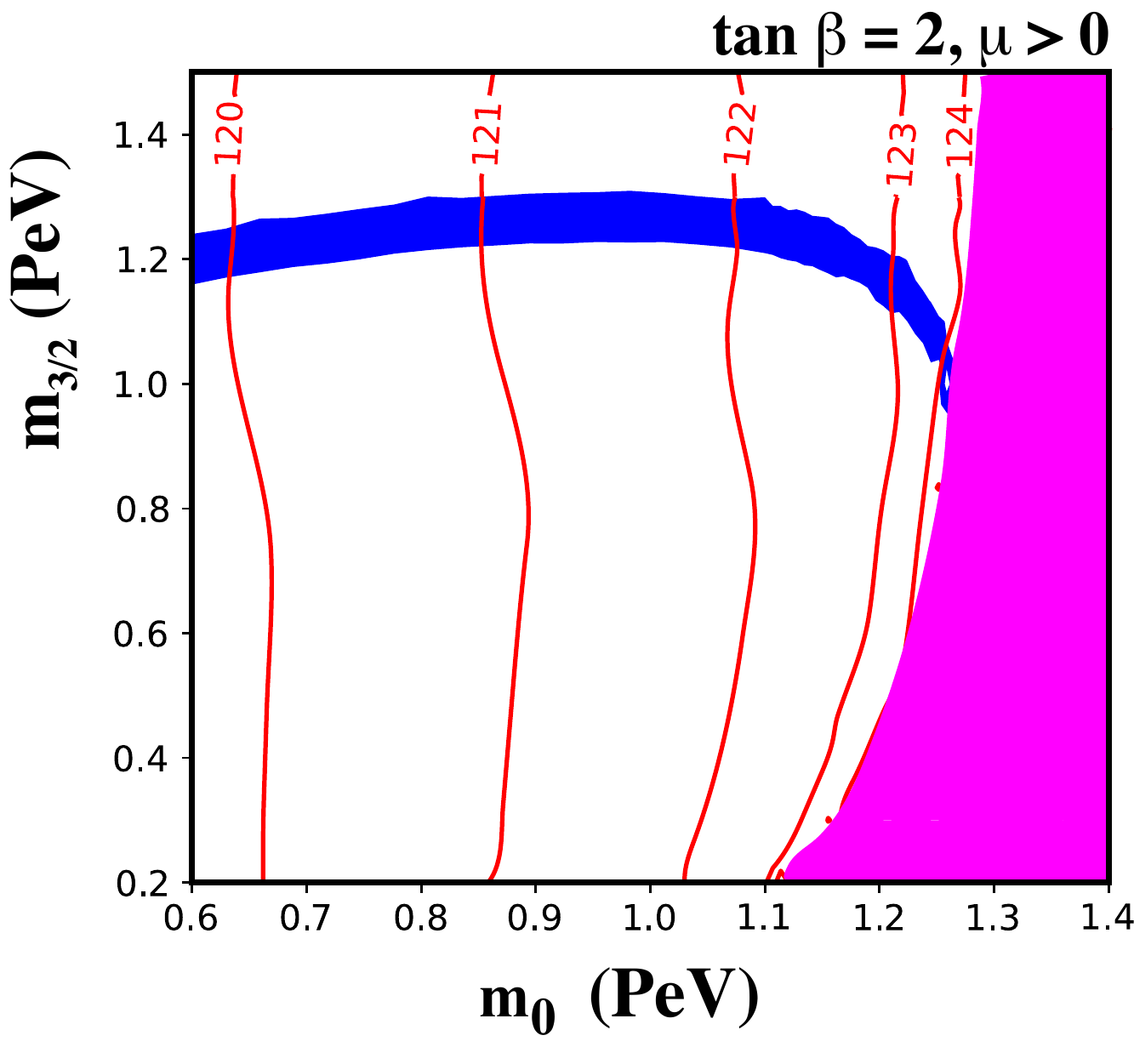}
\includegraphics[width=0.48\columnwidth]{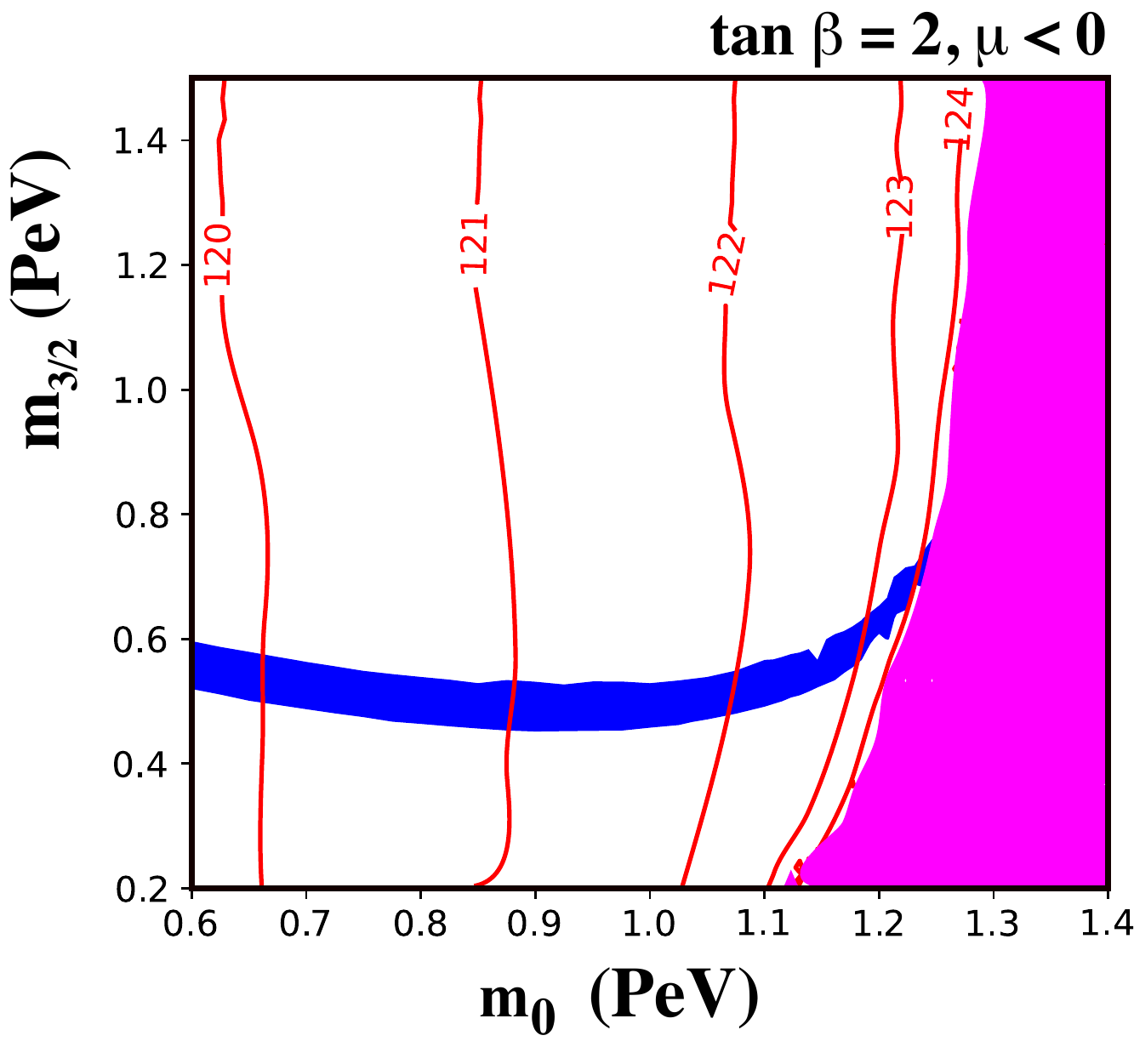} \caption{\it
The $(m_0, m_{3/2})$ plane in the mAMSB/PGM models for $\tan \beta = 5$ (upper panels) and $\tan \beta = 2$ (lower panels) for $\mu > 0$ (left panels) and $\mu < 0$ (right panels).
Electroweak symmetry breaking does not occur in the region shaded pink.
Contours of $m_h$ are shown as red lines labeled in units of GeV. 
The LSP relic density is $\Omega_\chi h^2 = 0.12 \pm 0.01$ in the region shaded blue.} 
 \label{fig:ehow++}
\end{figure}

As discussed earlier, the mAMSB model is characterized by three continuous parameters, $m_{3/2}, m_0$ and $\tan \beta$, and the sign of $\mu$. Examples of $(m_0, m_{3/2})$ planes with $\tan \beta = 5$ and 2 and both signs of $\mu$ are shown in Fig.~\ref{fig:ehow++}, which updates a similar plot given in \cite{ehow++}. In the pink shaded region to the right of each panel, there are no solutions for the minimization of the Higgs potential, and therefore radiative electroweak symmetry breaking is not possible in this region.
The red lines are contours of $m_h$ in GeV. 
(Our calculation of the Higgs mass is discussed in more detail below.)
In the largely horizontal blue-shaded region the LSP relic density is
$\Omega_\chi h^2 = 0.12 \pm 0.01$.

The wino mass increases monotonically with $m_{3/2}$ as seen in Eq.~(\ref{eq:M2}). For $\tan \beta = 5$ and $m_{3/2} \approx 0.9 - 1$~PeV, the wino mass is roughly 3 TeV and, as seen in the upper panels of Fig.~\ref{fig:ehow++}, it is able to provide the correct relic density when the Sommerfeld enhancements are included.
For relatively low values of $\tan \beta$ such
as those chosen in Fig.~\ref{fig:ehow++}, the Higgs mass is quite sensitive to parameter choices and increases with $\tan \beta$,
and we find that for $\tan \beta \simeq 5$
the relic density is satisfied for an
acceptable value of the Higgs mass. 

We note that, as $m_0$ increases, the value of the $\mu$ eventually starts to decrease so that the LSP becomes more Higgsino-like close to the region with no electroweak symmetry breaking. When $\mu \sim 1$ TeV and $m_{3/2}$ is sufficiently large the LSP is almost a pure Higgsino and the relic density
is acceptable. This region is visible as
the diagonal blue strip running close to the boundary of the pink-shaded region in the upper panels of Fig.~\ref{fig:ehow++}. However, we do not consider this strip, as our approximations for the elastic scattering cross section 
only apply in the wino-like case represented by the horizontal blue-shaded band \footnote{Electroweak corrections for the Higgsino-like LSP are found to be negligibly small~\cite{Hisano:2011cs, Hill:2011be, Hisano:2015rsa, Nagata:2014wma}.  }. 

As one can see in the upper panels of Fig.~\ref{fig:ehow++}, for $\tan \beta = 5$ the requirements of electroweak symmetry breaking and the cold dark matter density enforce $m_0 \ll m_{3/2}$. PGM boundary conditions with $m_0 \simeq m_{3/2}$ are only possible at lower values of $\tan \beta \simeq 2$ \cite{eioy}. In the lower two panels of  Fig.~\ref{fig:ehow++}, we show the $(m_0, m_{3/2})$ plane with $\tan \beta =2$. In the horizontal blue-shaded region, the LSP is again wino-like with the required relic density. The Higgs mass is generally too low, except for the largest values of $m_0$ allowed by radiative electroweak symmetry breaking. We note that higher values of $m_h$ can be attained for slightly higher values of $\tan \beta$. For example, $m_h = 125$ GeV is possible for $\tan \beta = 2.05$.  

\begin{figure}[!ht]
  \centering
\includegraphics[width=0.47\columnwidth]{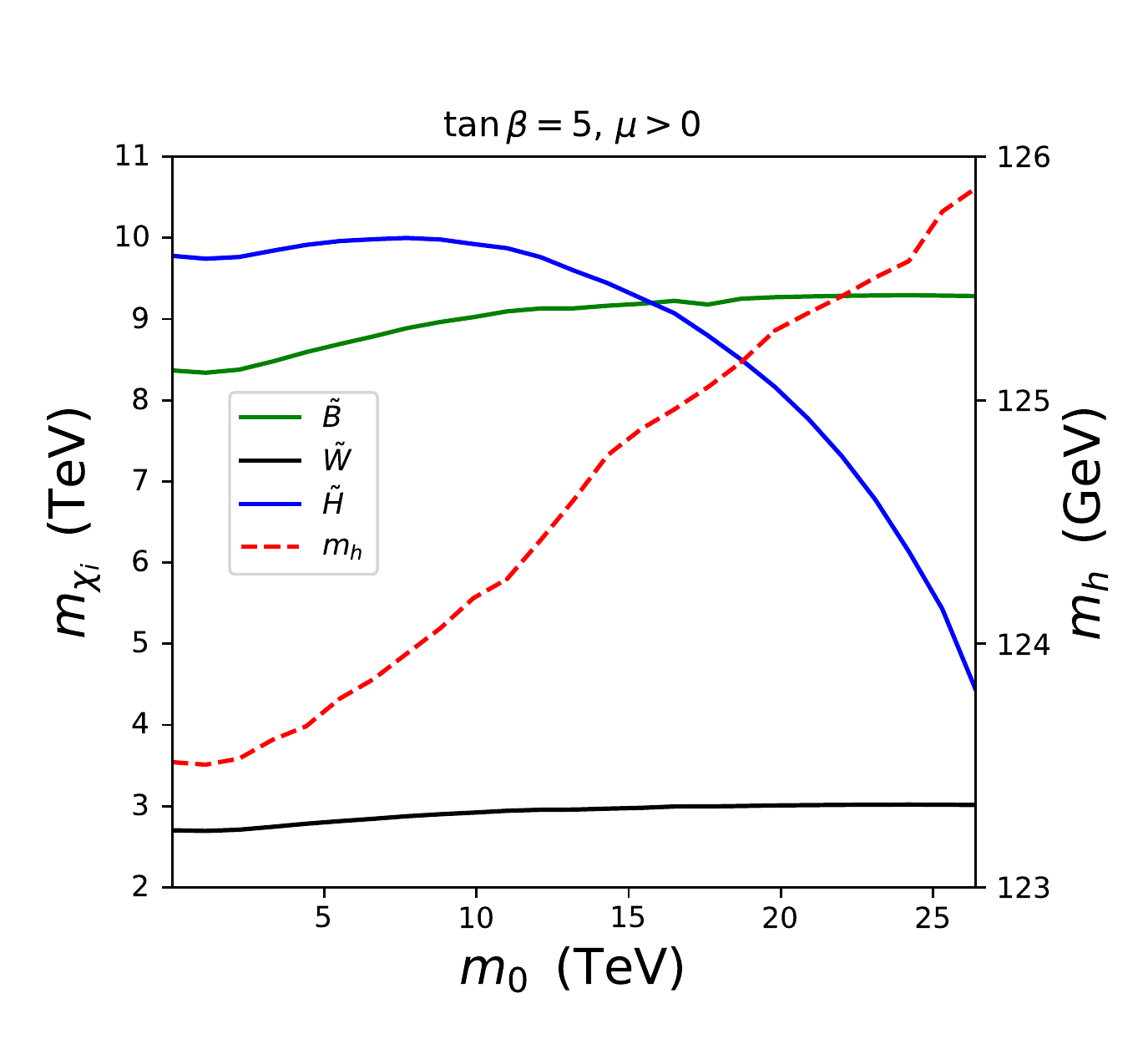}
\includegraphics[width=0.47\columnwidth]{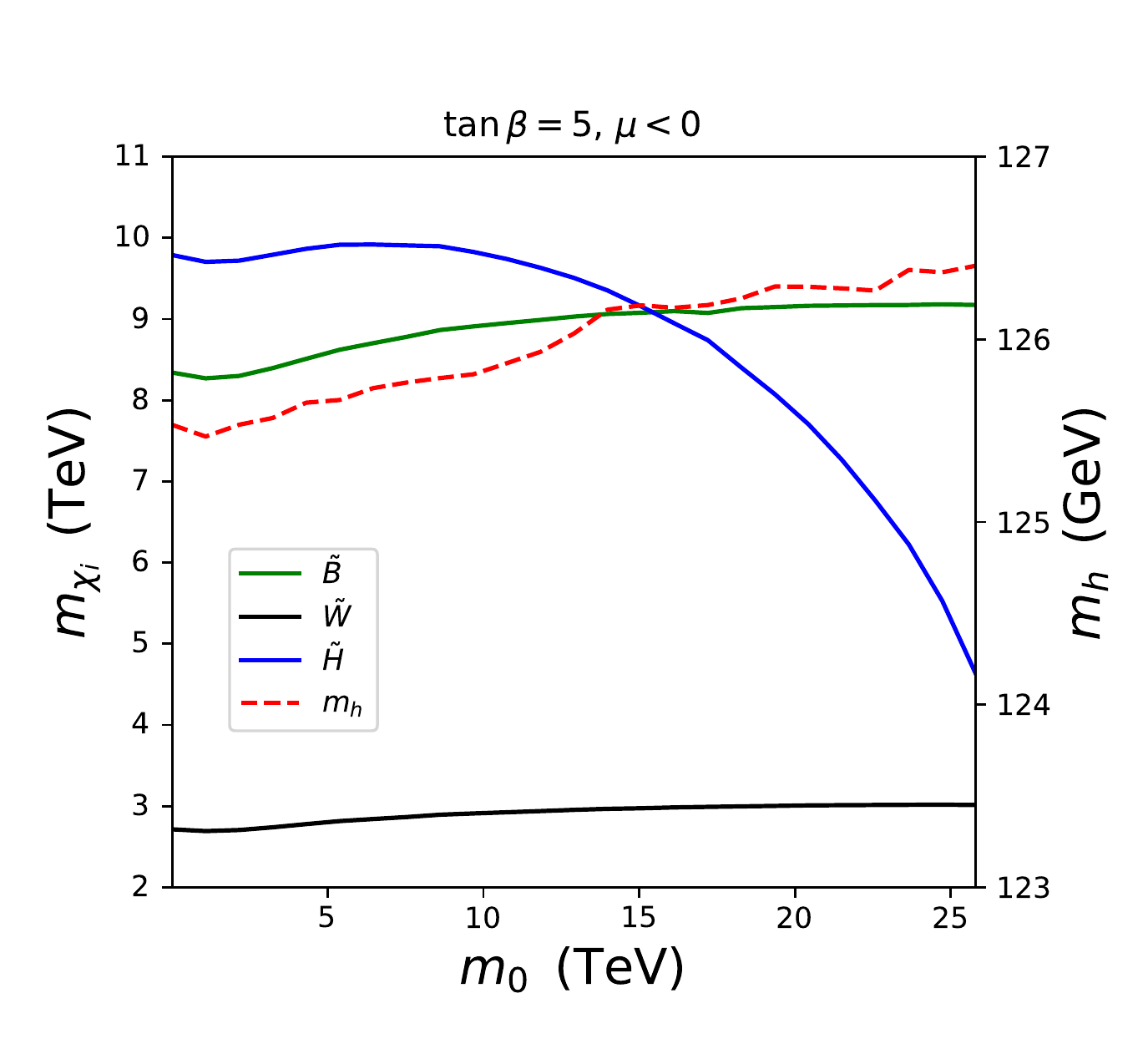} \\
\includegraphics[width=0.47\columnwidth]{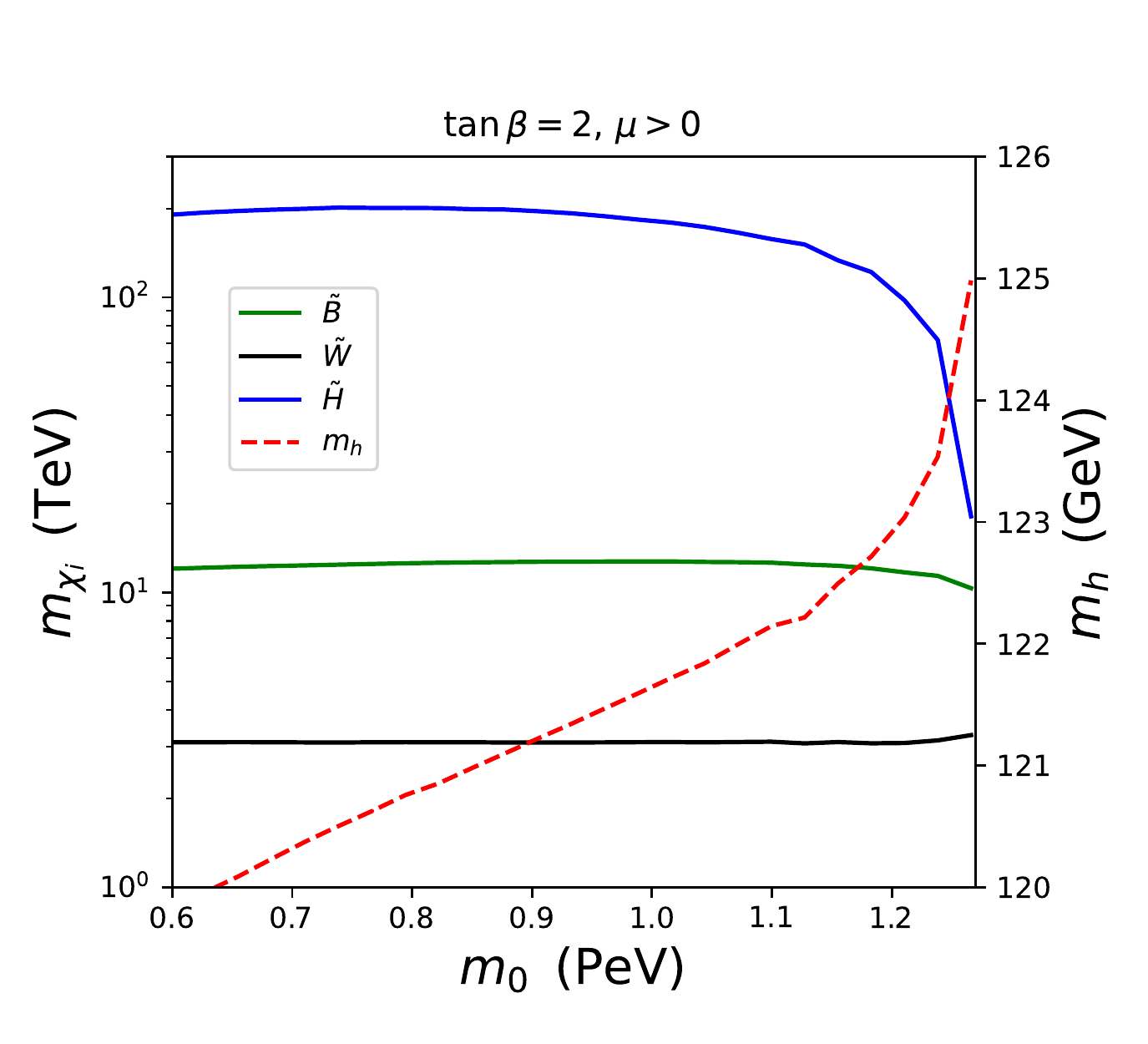}
\includegraphics[width=0.47\columnwidth]{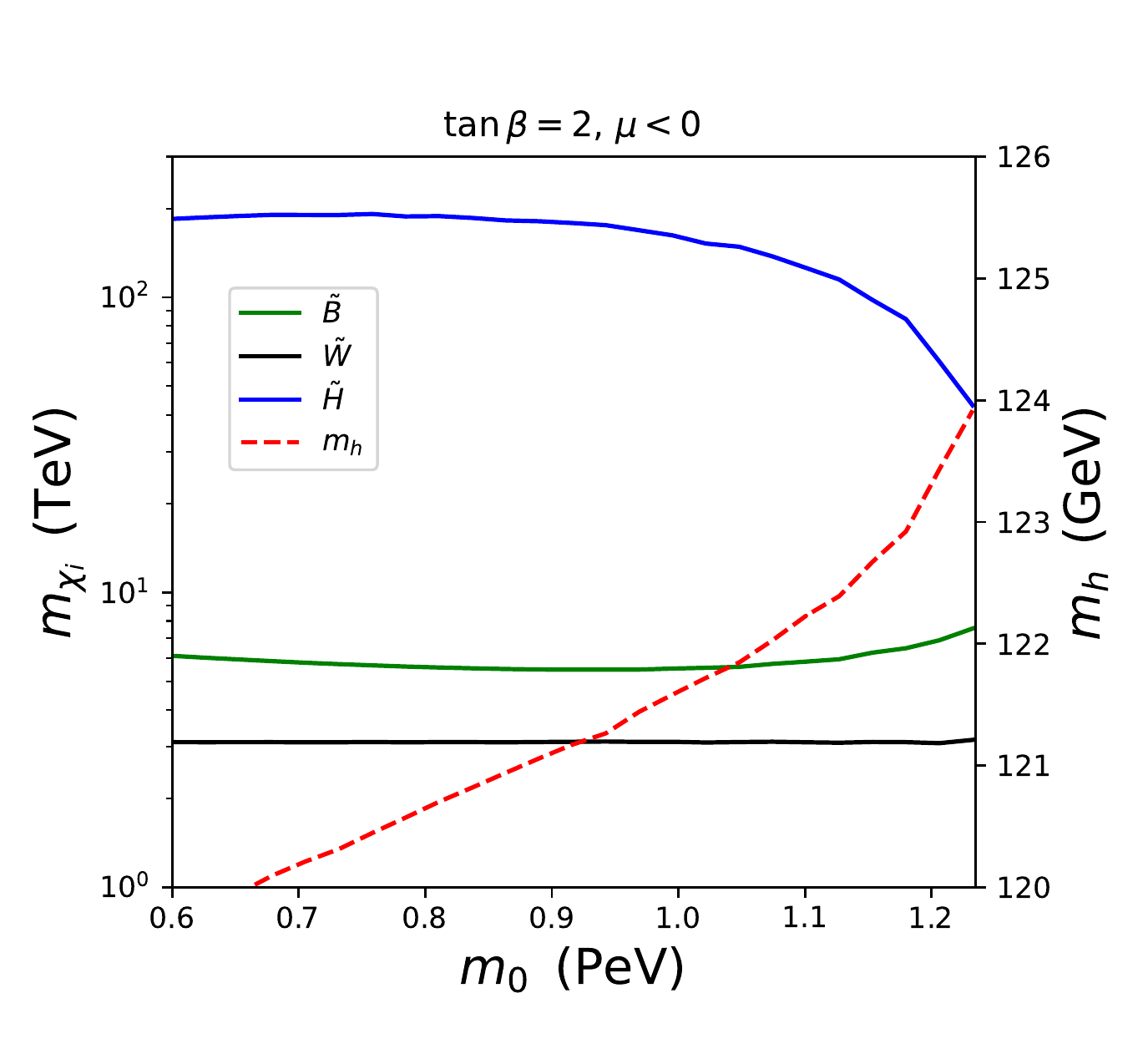}
  \caption{\it 
  Calculations of the neutralino and Higgs masses as functions of $m_0$ for $\tan \beta = 5$ (upper panels) and $\tan \beta = 2$ (lower panels) and $\mu > 0$ (left panels)
  and $\mu < 0$ (right panels). 
  The (Higgsino-) (bino-) wino-like mass contours are (blue) (red) black. For each value of $m_0$, the value of $m_{3/2}$ is chosen so that the relic LSP density is $\Omega_\chi h^2 = 0.12$.  Also shown is the Higgs mass given by the scale on the right.} 
  \label{fig:mchimap}
\end{figure}

The neutralino mass spectra for $\tan \beta = 2$ and 5 are shown in Fig.~\ref{fig:mchimap} for both signs of $\mu$.
Along each of the curves, the value of $m_{3/2}$ is chosen so that the relic density is $\Omega_\chi h^2 = 0.12$.  For $\tan \beta = 5$, the range of gravitino masses is 0.87--0.95 PeV for $\mu > 0$ and 0.86--0.94 PeV for $\mu < 0$. Similarly, the range for the gravitino mass when $\tan \beta = 2$ is  1.0--1.3 PeV for $\mu > 0$ and 0.49--0.70 PeV for $\mu < 0$. In all of the cases shown, we see that both the bino and wino masses are relatively independent of $m_0$ when $m_0 \ll m_{3/2}$.
On the other hand, the Higgsino mass is quite sensitive to $m_0$, as it is essentially determined by $\mu$, which is in turn fixed by the electroweak symmetry breaking conditions. At very large $m_0$,
$|\mu|$ begins to drop, and at sufficiently large $m_0$
the focus-point region~\cite{fp} is reached and the Higgsino becomes the LSP.

\begin{figure}[!ht]
  \centering
\includegraphics[width=0.47\columnwidth]{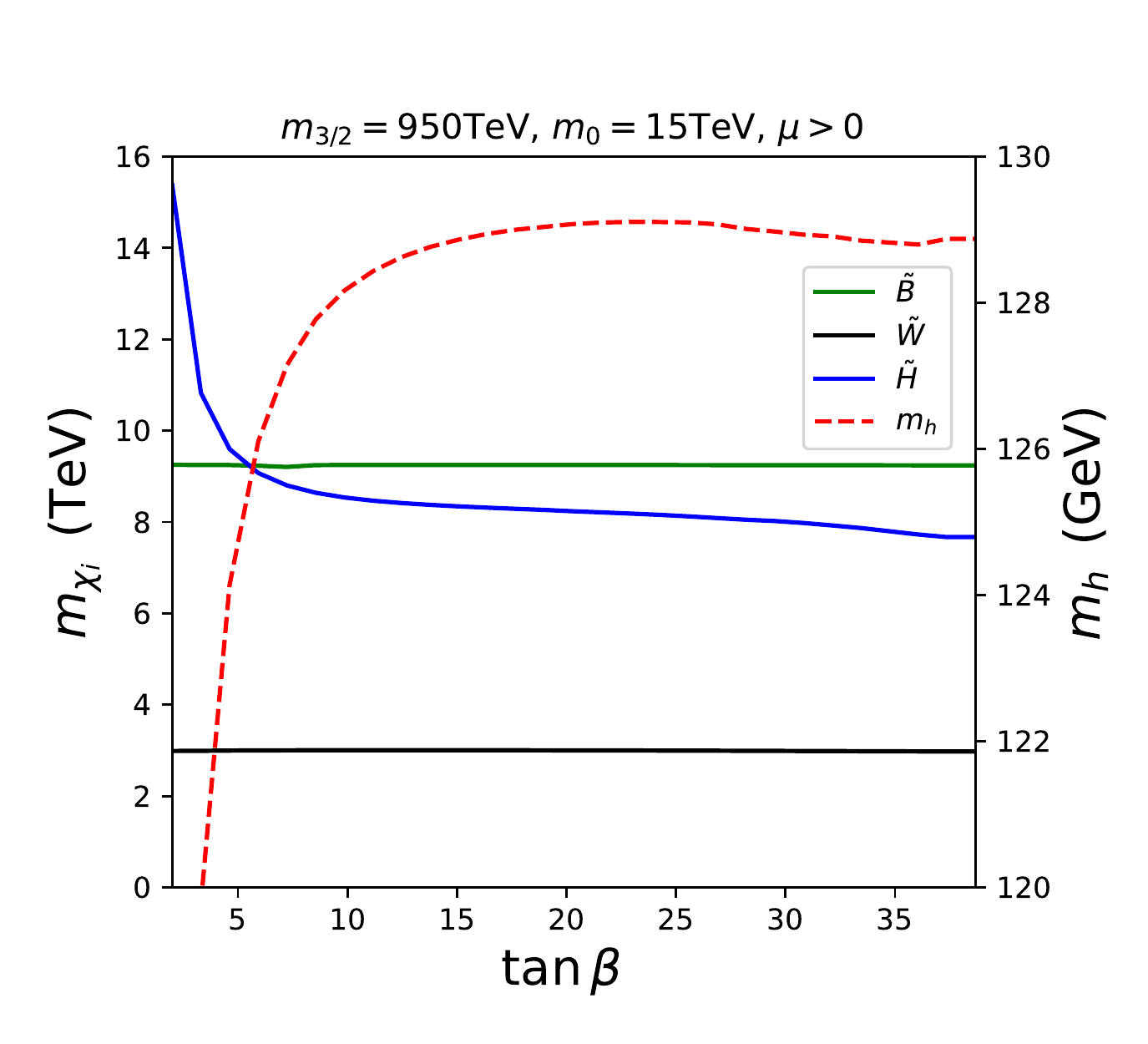}
\includegraphics[width=0.47\columnwidth]{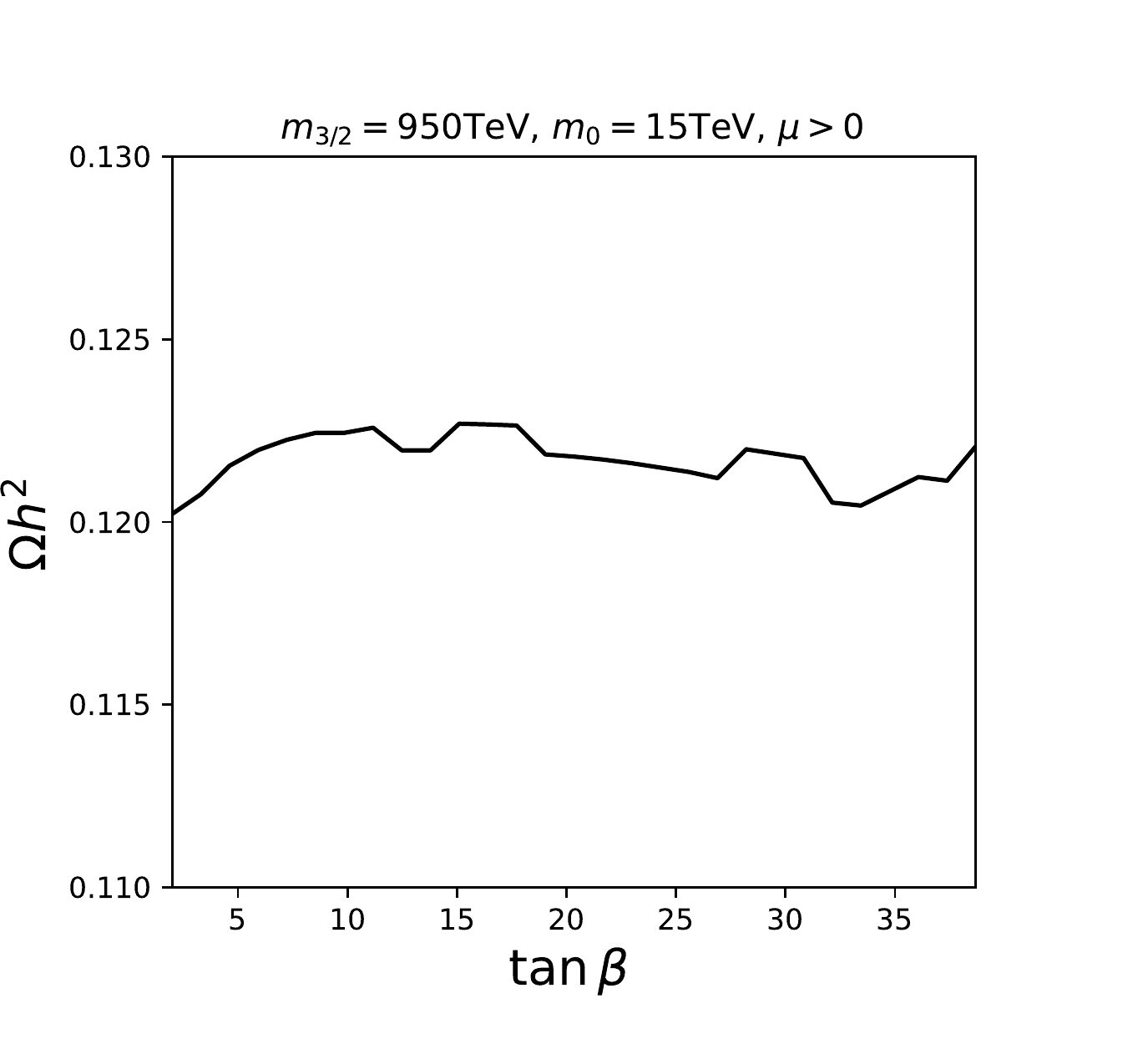}\\
 \includegraphics[width=0.47\columnwidth]{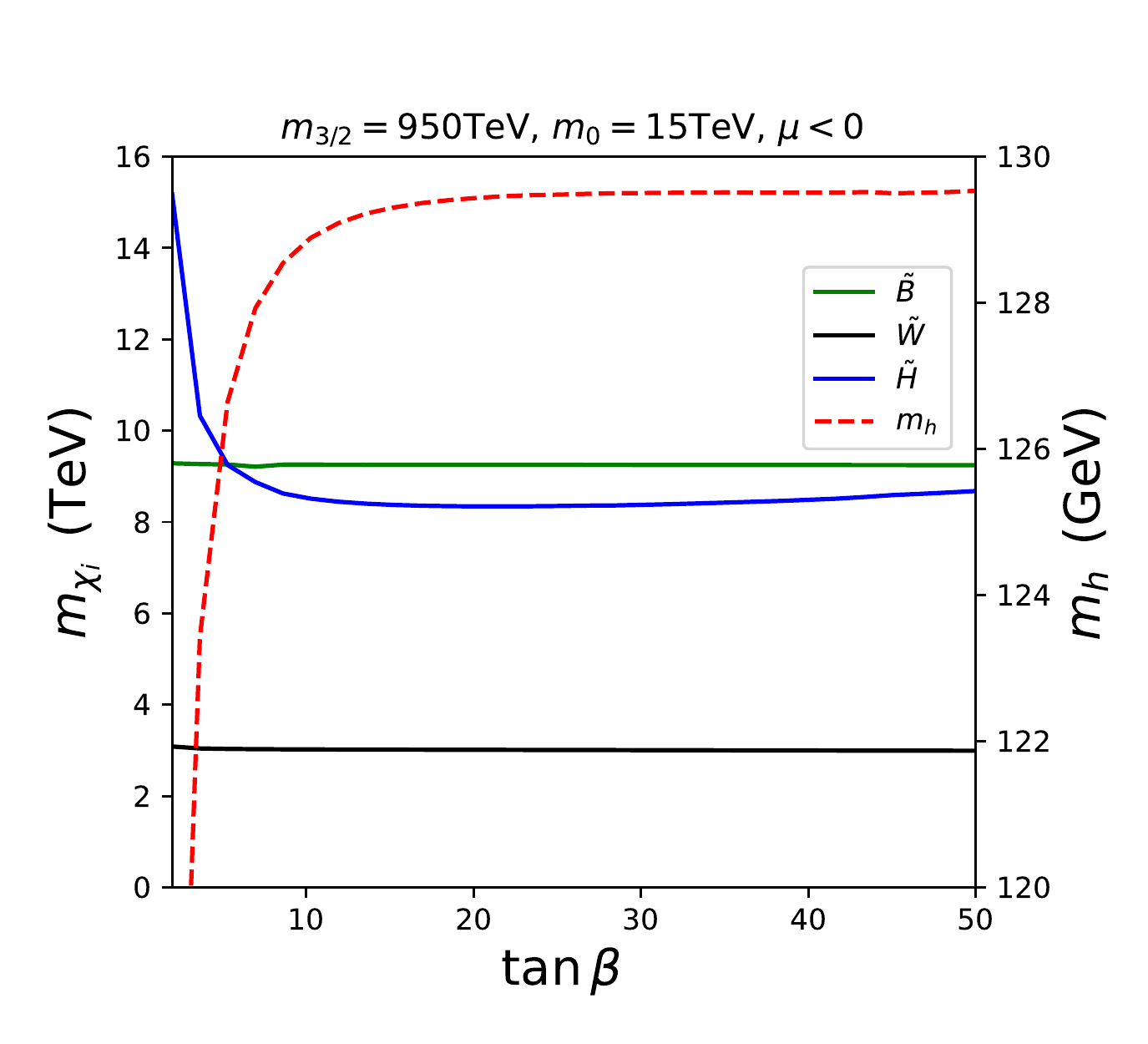}
 \includegraphics[width=0.47\columnwidth]{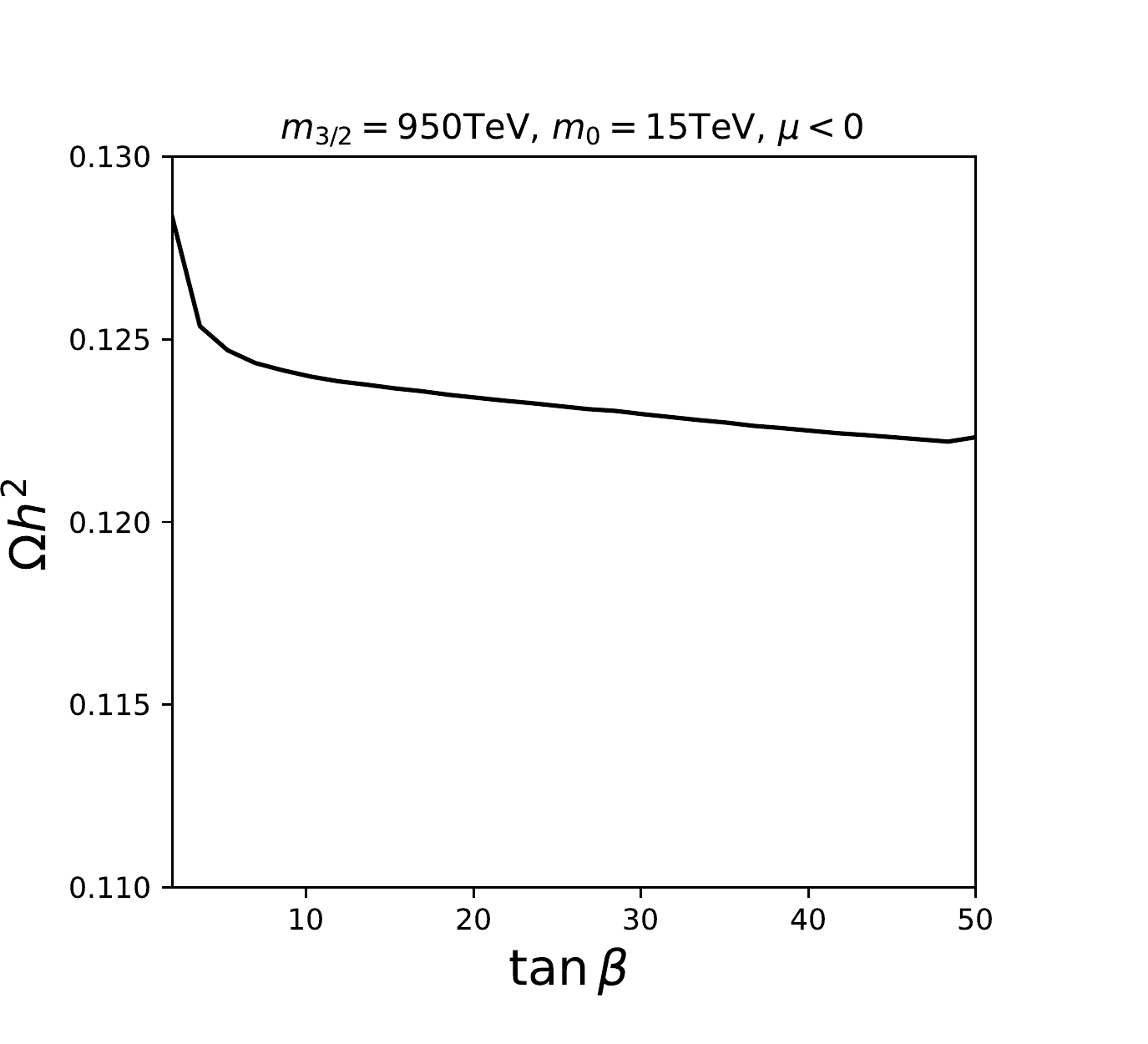}
  \caption{\it Neutralino and Higgs masses (left panels) and the cold dark matter density (right panels) as functions of $\tan \beta$ for $m_{3/2} = 950$ TeV, $m_0 = 15$ TeV with $\mu > 0$ (upper panels) and $\mu < 0$ (lower panels).
The lightest neutralino is always wino-like (black line). The bino-like 
neutralino mass is shown as a green line and the Higgsino-like neutralino
masses as a blue line. (The two Higgsinos are nearly degenerate.) The red dashed line shows the Higgs mass.}
  \label{fig:mchitbmap}
\end{figure}

It is instructive to consider the dependences of the neutralino masses on 
$\tan \beta$ as shown in the left panels of Fig.~\ref{fig:mchitbmap} for the cases $m_{3/2} = 950$~TeV,
$m_0 = 15$~TeV and both signs of $\mu$. We see that the lightest 
neutralino is always wino-like with a mass around 3~TeV, and the bino-like neutralino
has a mass $\sim 9$~TeV. The second lightest
neutralino is bino-like for $\tan \beta \lesssim 5$ and Higgsino-like for larger
$\tan \beta$. Also shown in the left panels of Fig.~\ref{fig:mchitbmap} is the Higgs mass plotted as a function of $\tan \beta$. Here we see clearly the strong dependence of $m_h$ on $\tan \beta$ for $\tan \beta \lesssim 10$. All of the masses are relatively independent of $\tan \beta$ when $\tan \beta \gtrsim 10$.

For the same choices of parameters, we show the relic LSP density as a function of $\tan \beta$ in the right panels of Fig.~\ref{fig:mchitbmap}.
While it might appear that there is a strong dependence on $\tan \beta$, this is largely due to the choice of scale. The relic density is acceptable (particularly when calculational uncertainties are taken into account) throughout the range of $\tan \beta$ shown.

\begin{figure}
	\centering
 	\includegraphics[width=0.45\columnwidth]{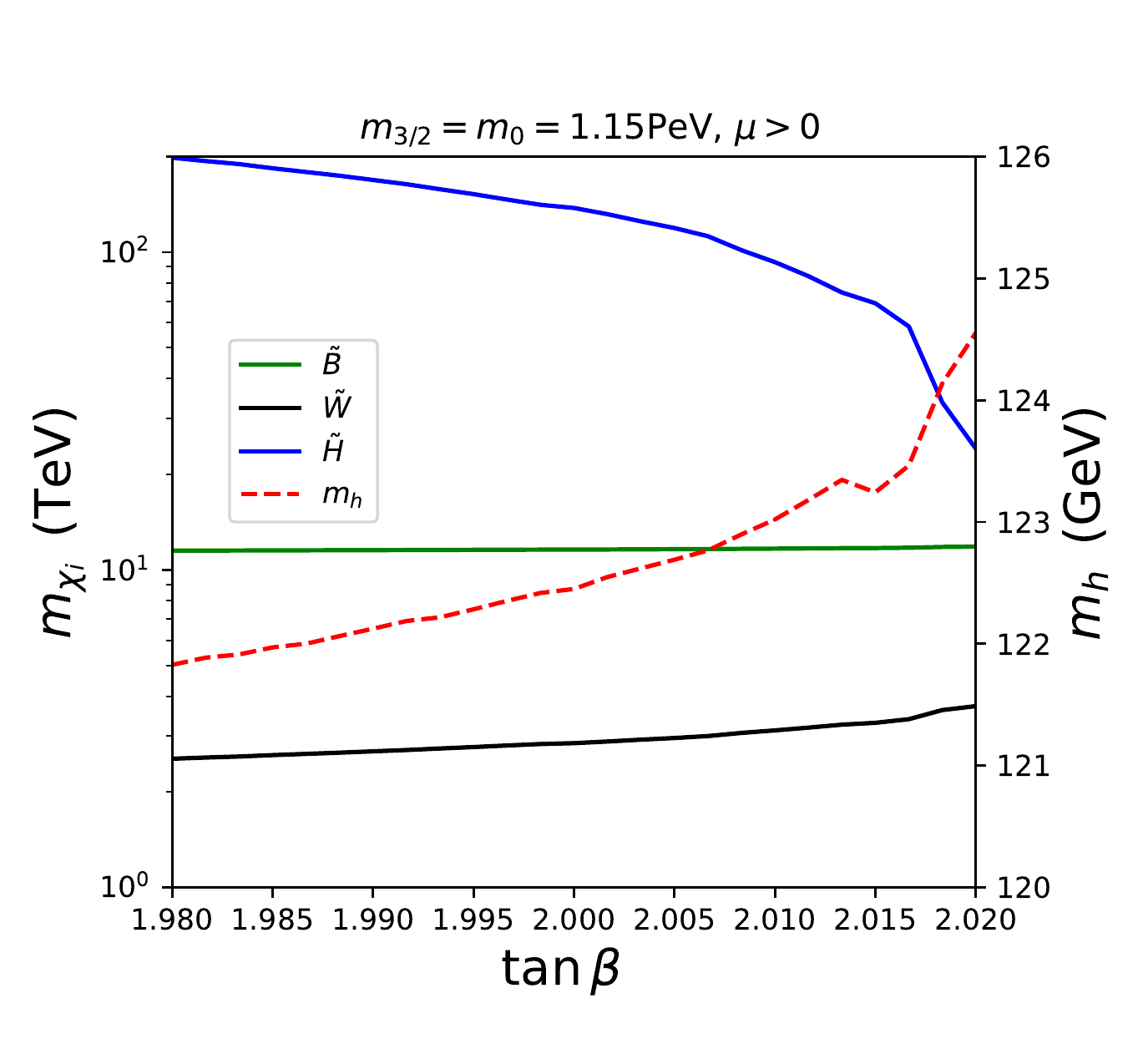}
	\includegraphics[width=0.47\columnwidth]{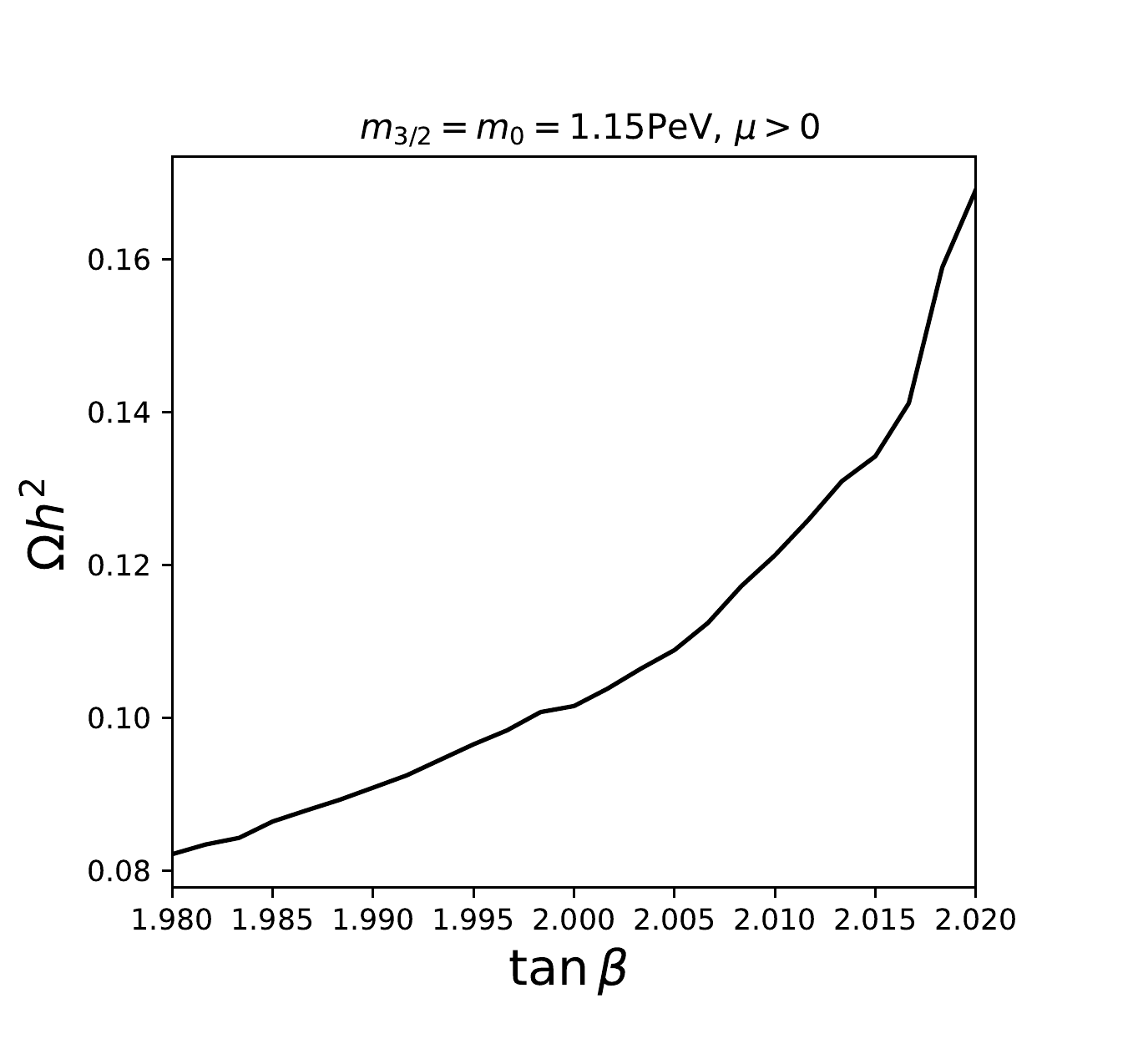}\\
 	\includegraphics[width=0.45\columnwidth]{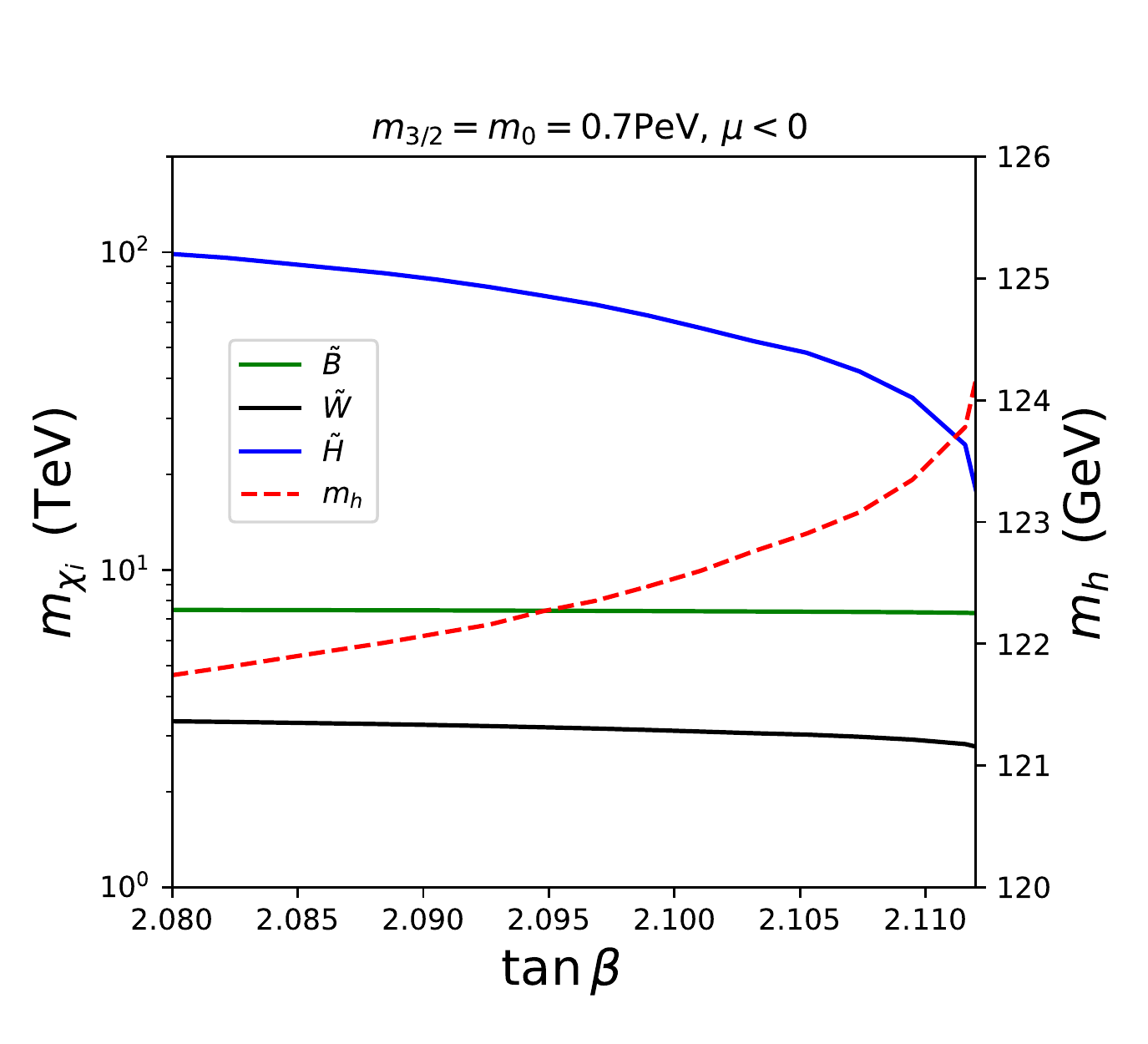}
    \includegraphics[width=0.47\columnwidth]{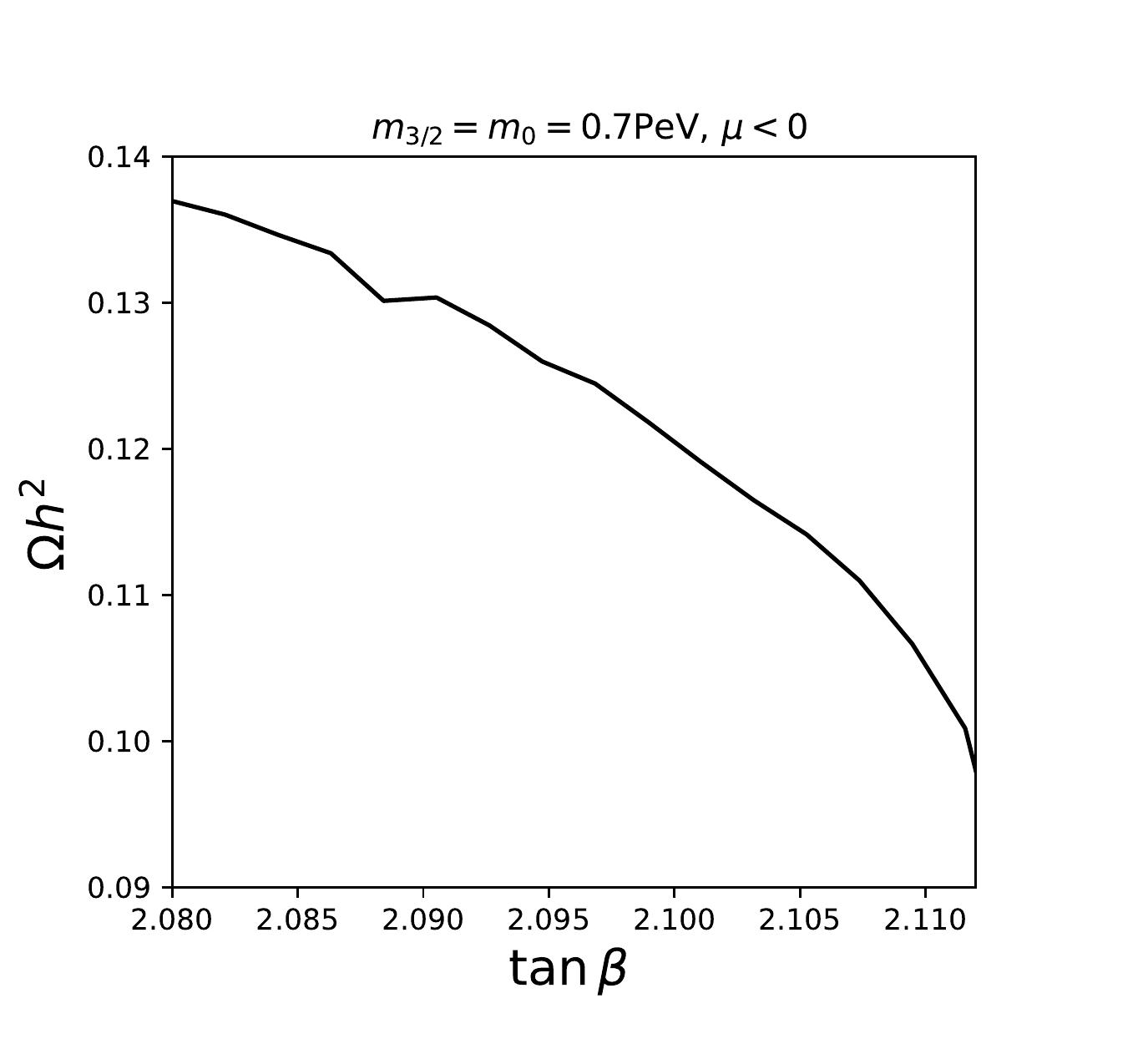}
	\caption{\it Neutralino and Higgs masses (left panels) and the relic LSP density (right panels) as functions of $\tan \beta$ for the PGM model when $m_{3/2} = m_0 = 1.15$~PeV with $\mu>0$ (upper panels) and $\mu < 0$ (lower panels). } 
  \label{pgm}
\end{figure}

In the limit that $m_0 = m_{3/2}$ as in PGM models,
only a relatively narrow range of $\tan \beta$ is allowed,
as seen in Figs.~\ref{pgm} where the neutralino and Higgs masses and relic density are plotted as functions of $\tan \beta$ assuming $m_0 = m_{3/2} = 1.15$ PeV for $\mu > 0$ (upper panels) and $m_0 = m_{3/2} = 0.7$ PeV for $\mu < 0$ (lower panels). The mass spectra plotted in the left panels show again a wino-like LSP with mass around 3 TeV. Because the $\mu$ parameter typically takes values of order $m_0$, the Higgsino masses are very large in this case until $\tan \beta$ is sufficiently large that the focus-point region is approached. At this point, the Higgs mass is increased toward its experimental value. At still larger $\tan \beta$ radiative electroweak symmetry breaking is no longer possible.

The relic LSP density for the same sets of parameters is shown in right panels of Fig.~\ref{pgm}. Depending on the sign of $\mu$, the relic LSP density (which is always close to the observationally determined values) increases ($\mu > 0$) or decreases ($\mu < 0$), tracking the behavior of the wino-like LSP mass. 

\subsection{Calculation of the Higgs Boson Mass}

\begin{figure}[!ht]
  \centering
  \subcaptionbox{\label{fig:mh0} $\mu > 0$}{
  \includegraphics[width=0.48\columnwidth]{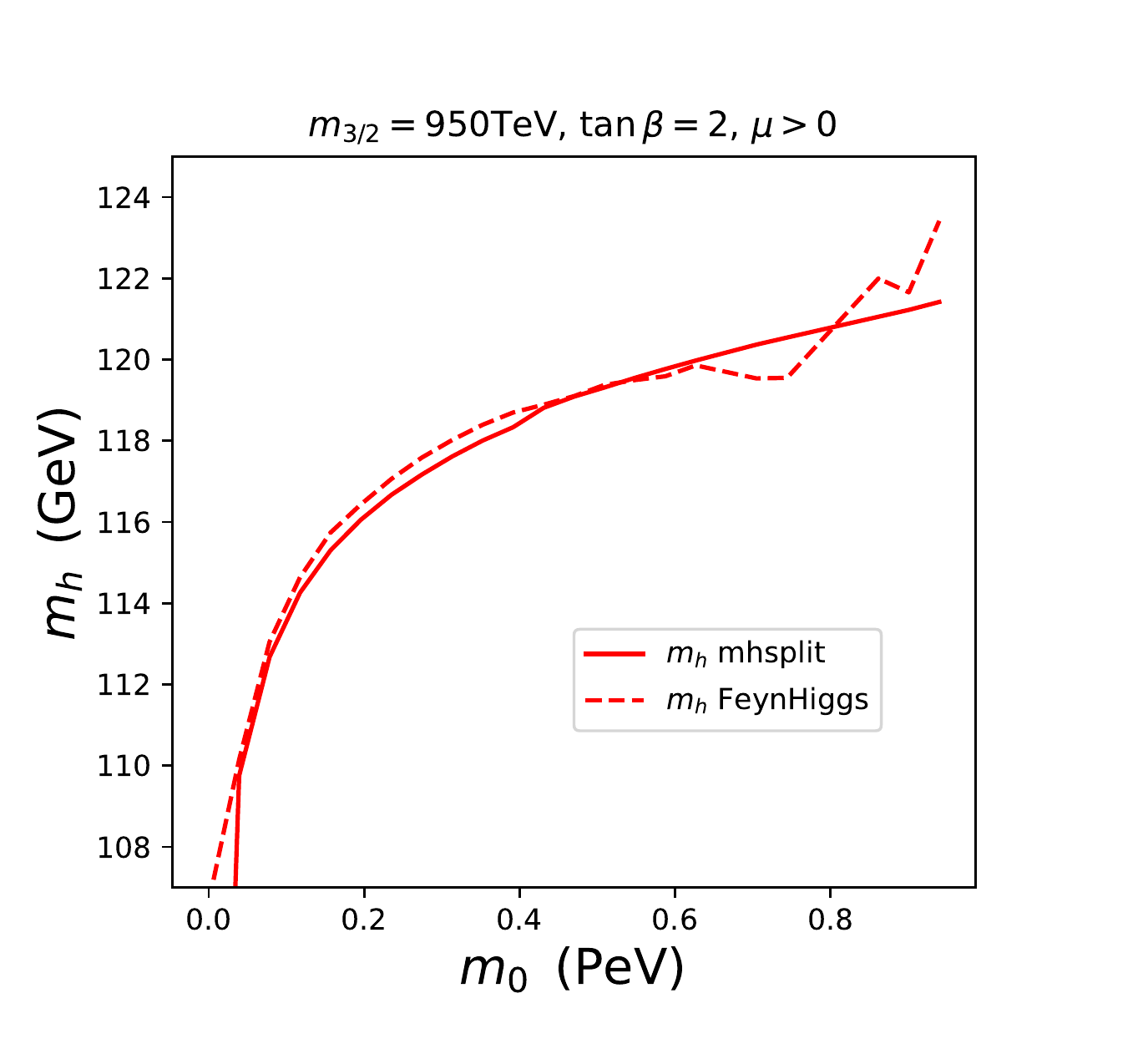}}
  \subcaptionbox{\label{fig:mh1} $\mu < 0$}{
  \includegraphics[width=0.48\columnwidth]{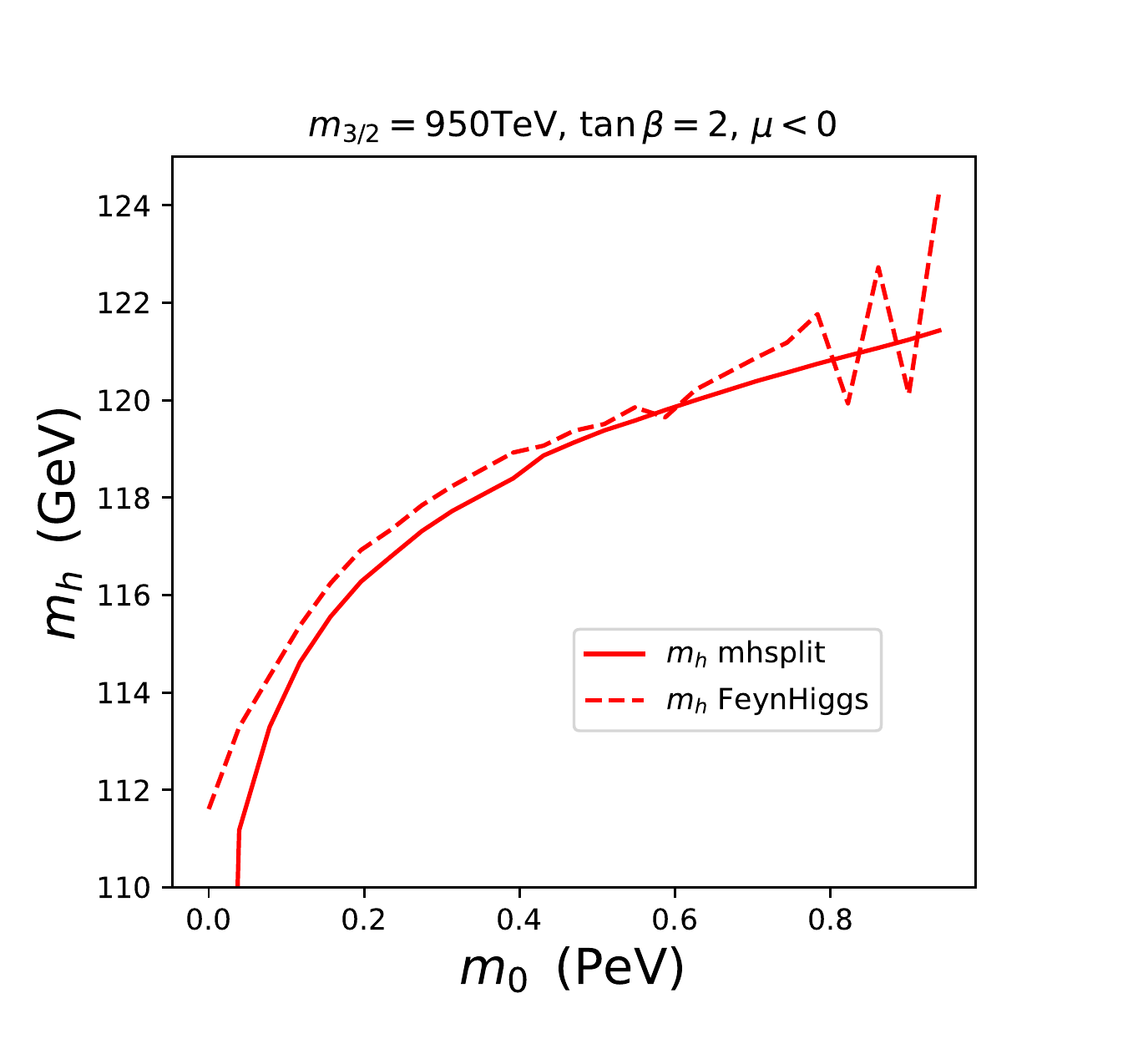}}
\caption{\it  Comparisons of the calculations of $m_h$
made using {\tt FeynHiggs~2.18.1} (dashed lines) and {\tt mhsplit},
a simplified
code devised specifically for large supersymmetry-breaking scales
(solid lines). We see that the simplified calculation of $m_h$ is
very stable at large values of $m_0$ for $m_{3/2} = 950$~TeV, 
$\tan \beta =2$ and both signs of $\mu$, and agrees very well with the
{\tt FeynHiggs} results for values of $m_0 \simeq 0.1-0.6$~PeV.
} 
  \label{fig:mh}
\end{figure}

Reliable calculations of the light Higgs mass are challenging when
the supersymmetry-breaking scale is ${\cal O}({\rm PeV})$, as is the
case for the AMSB and PGM models considered here. Accordingly, we have
compared results from two codes: {\tt FeynHiggs~2.18.1} \cite{FH} and {\tt mhsplit}, a simplified
code devised specifically for large supersymmetry-breaking scales. Following \cite{mhsplit,gs}, the 2-loop RGE evolution of the couplings of the effective theory below the supersymmetry breaking scale are used to obtain the Higgs mass at full next-to-leading order accuracy as described in more detail in  \cite{Dudas:2012wi}
for a model with an AMSB-like spectrum. The same code was also adapted to PGM models \cite{eioy}.
A comparison of the results of the two codes is shown in Fig.~\ref{fig:mh}, where we plot the Higgs mass as a function of $m_0$ given by each code for $m_{3/2} = 950$ TeV and $\tan \beta = 2$ with $\mu > 0$ (left) and $\mu < 0$ (right). The Higgs mass from {\tt mhsplit} is shown by the (smooth) solid curve, and that from {\tt FeynHiggs~2.18.1} by the dashed curve. For $m_0 = 0.1 - 0.6$ PeV, both codes are in good agreement and appear well-behaved. At lower values of $m_0$,
the approximations used in  {\tt mhsplit} break down. At $m_0 \gtrsim 0.6$ PeV, {\tt FeynHiggs~2.18.1} starts to exhibit irregularities that are absent in
the simplified calculation.

Motivated by the above comparison, the Higgs mass contours for the mAMSB models shown in the upper panels of Fig.~\ref{fig:ehow++} are run using  {\tt FeynHiggs~2.18.1}, whereas those in the lower panels for the PGM-like models with significantly higher supersymmetry-breaking mass parameters are run using
{\tt mhsplit}. 
The Higgs mass as a function of $m_0$ for fixed $\tan \beta$ and $m_{3/2}$ fixed to yield $\Omega_\chi h^2 = 0.12$ is shown in Fig.~\ref{fig:mchimap}.
The upper panels with $\tan \beta = 5$ are calculated using {\tt FeynHiggs~2.18.1} while the lower panels with 
$\tan \beta = 2$ are calculated using {\tt mhsplit}.
As one can see, the scalar mass range there is highly sensitive to $\tan \beta$. It is relatively easy in the mAMSB models with $\tan \beta = 5$ to obtain acceptable Higgs masses with $m_0 \sim \mathcal{O}(10)$ TeV, whereas
$m_0 \gtrsim 1$ PeV is required in the PGM model with $\tan \beta = 2$, 
 and an acceptable Higgs mass is only possible when the Higgsino mass drops precipitously as one approaches the focus-point region. The dependence of the Higgs mass on $\tan \beta$ is shown in Fig.~\ref{fig:mchitbmap} for the mAMSB models and in the left panels of Fig.~\ref{pgm} for the PGM models.

\section{Results for the Spin-Independent Wino Scattering Cross Section}
\label{sec:results}

We now present our results for the spin-independent scattering cross section
of a wino-like neutralino, $\sigma_p$, obtained
using the results presented in Section~\ref{sec:spin-independentcalx}.
Explicit expressions for the mass functions introduced there are given in the Appendix.
We also provide in the Appendix simple approximate formulae that apply when the
wino-like neutralino mass $m_\chi$ is much larger than the electroweak scale, which
is the case for the parameter regions of interest in this paper.

\begin{figure}
  \centering
\includegraphics[width=0.48\columnwidth]{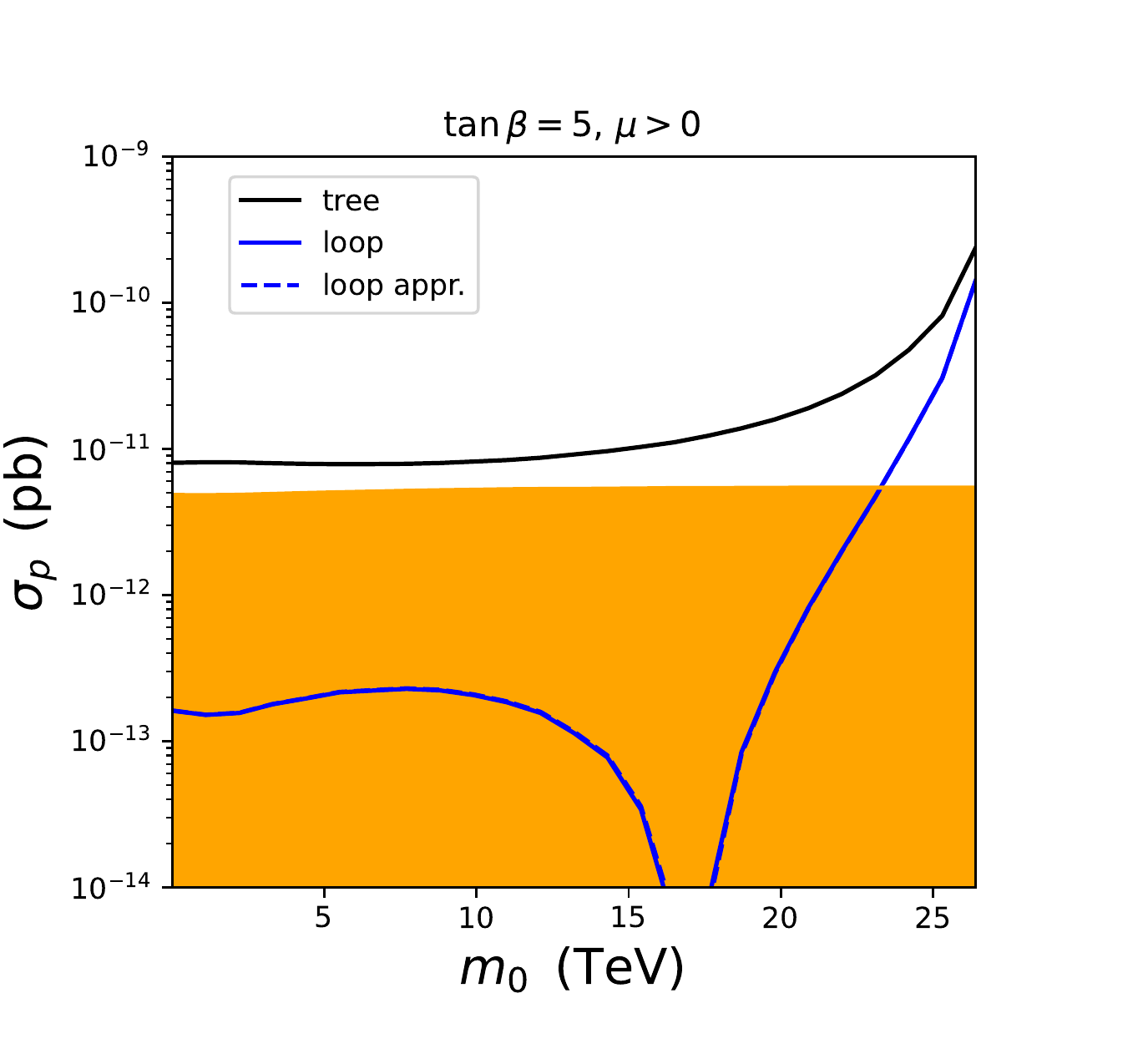}
\includegraphics[width=0.48\columnwidth]{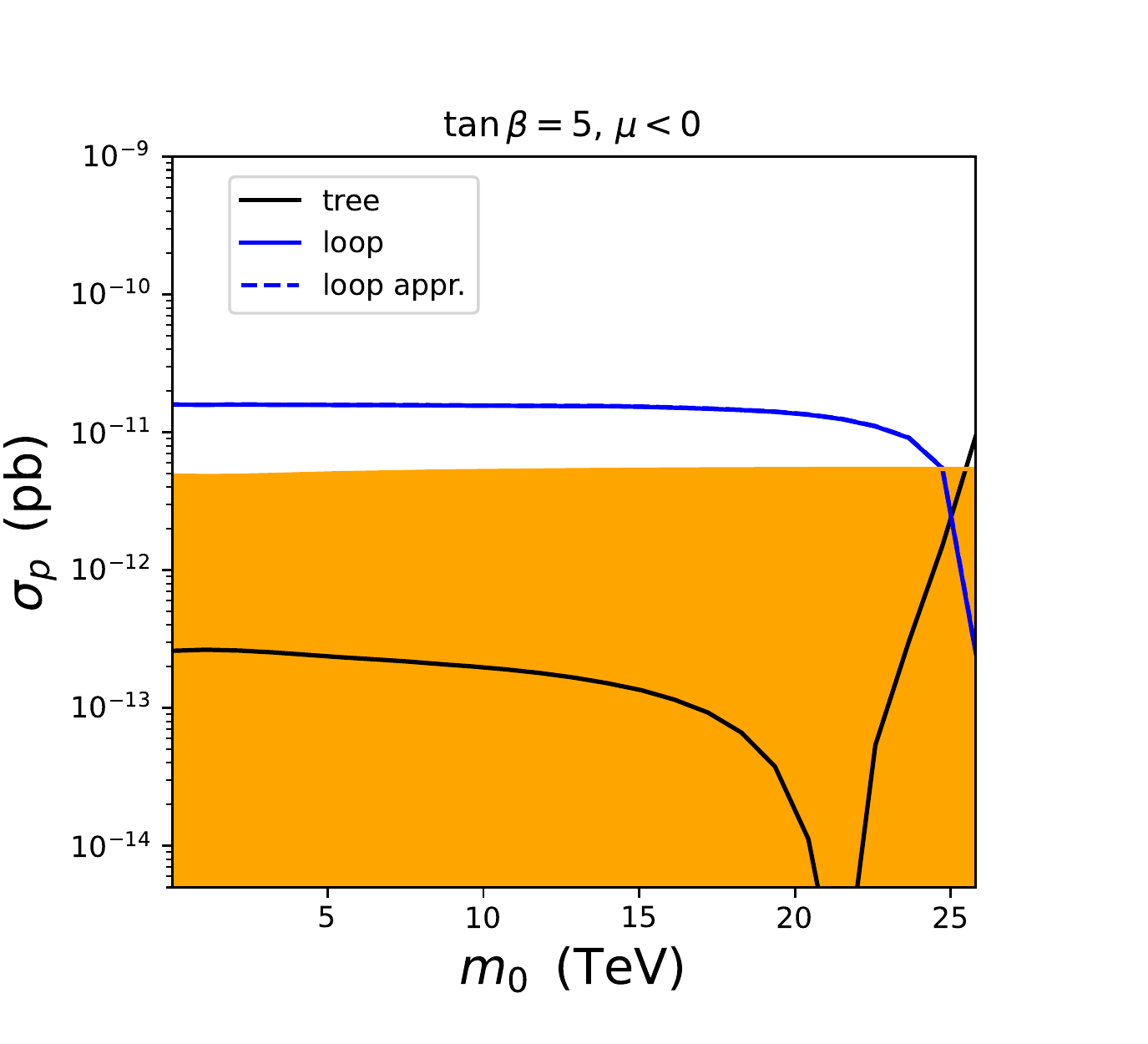} \\
\includegraphics[width=0.48\columnwidth]{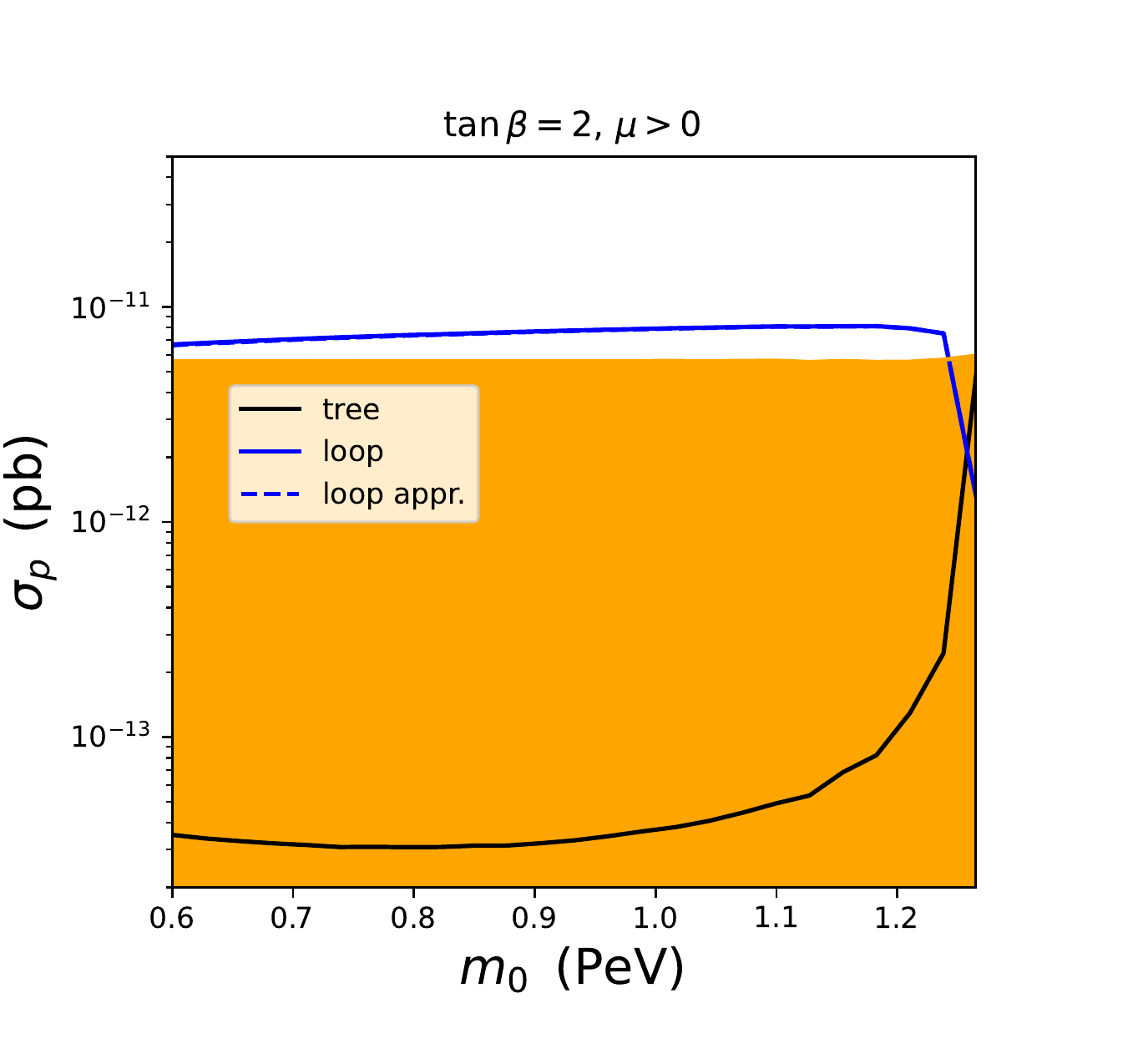}
\includegraphics[width=0.48\columnwidth]{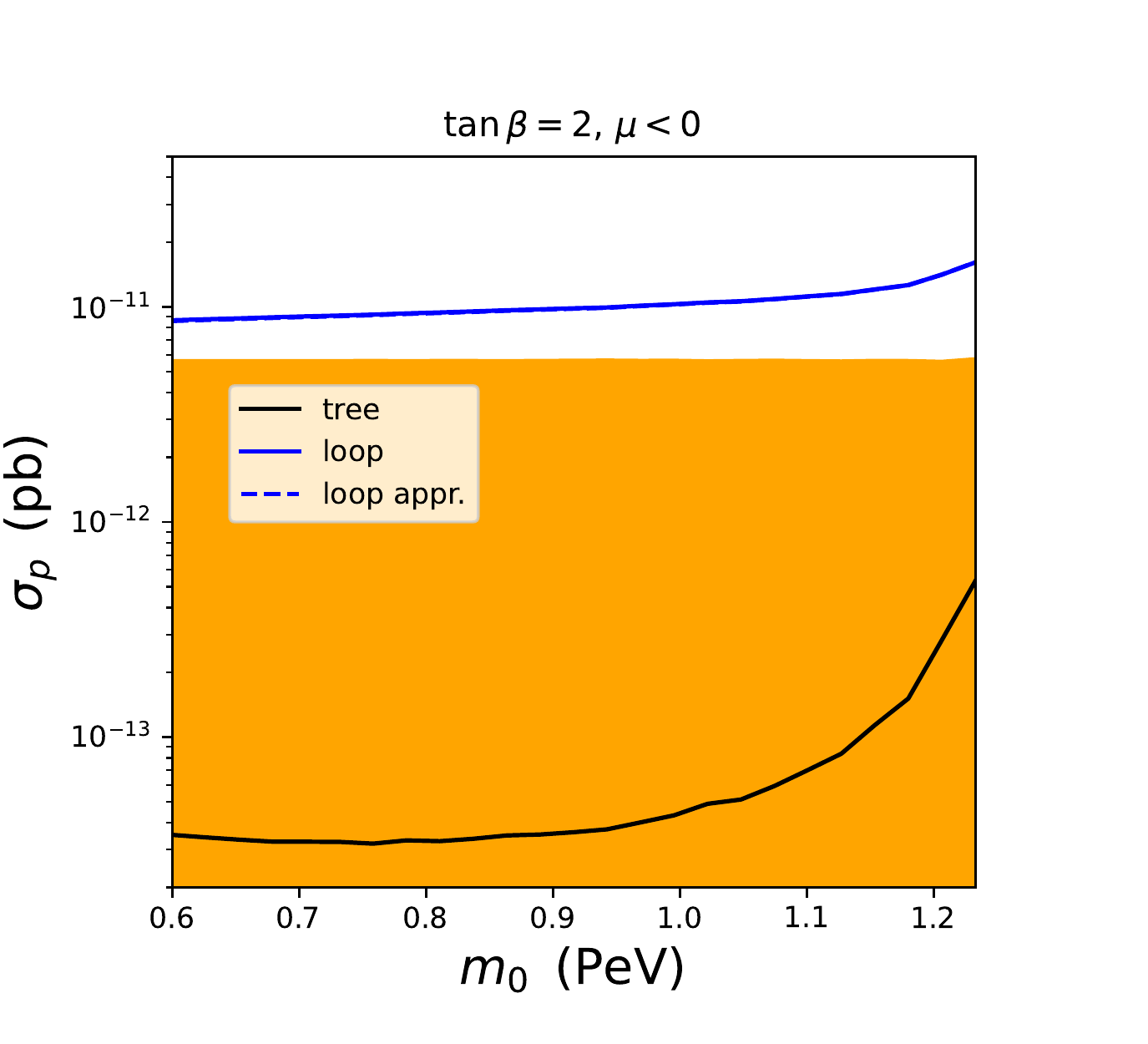}
  \caption{\it 
  Comparison of calculations of $\sigma_p$ as a function of $m_0$
made using tree-level matrix elements (black lines) and
one-loop calculations made using the full expressions in the Appendix
(solid blue lines) and the simplified expressions in Eq.~(\ref{simplified})
that become exact when the wino neutralino mass $m_\chi \gg m_W$
(dashed blue lines). The calculations are for the AMSB model with $m_{3/2} = 950$~TeV, $m_0 = 15$~TeV and both signs of $\mu$ (upper
panels) and the PGM-like model with $m_{3/2} = m_0 = 1.15$~PeV for $\mu > 0$ (lower left panel)
and with $m_{3/2} = m_0 = 0.7$~PeV for $\mu < 0$ (lower left panel). The orange-shaded region
corresponds to cross section values in the neutrino fog.} 
  \label{fig:mchisigmap}
\end{figure}

We present in Fig.~\ref{fig:mchisigmap} a comparison of calculations of $\sigma_p$
made using tree-level matrix elements (solid black lines) and
one-loop calculations (solid blue lines). The spin-independent cross section is plotted as a function of $m_0$ for the AMSB model with
$\tan \beta = 5$ (upper panels) for both $\mu > 0$ (left panels) 
and $\mu < 0$ (right panels). In the lower panels we show the cross section in the PGM-like model with $\tan \beta = 2$. The orange-shaded region in each plot corresponds to cross sections in the neutrino fog~\cite{floor}. 
For comparison, we note that for $m_\chi \simeq 3$ TeV the current best experimental limit, from the LUX-ZEPLIN (LZ) experiment~\cite{LZ}, is $\sigma_p \lesssim 10^{-9}$ pb.

\begin{figure}
  \centering
\includegraphics[width=0.48\columnwidth]{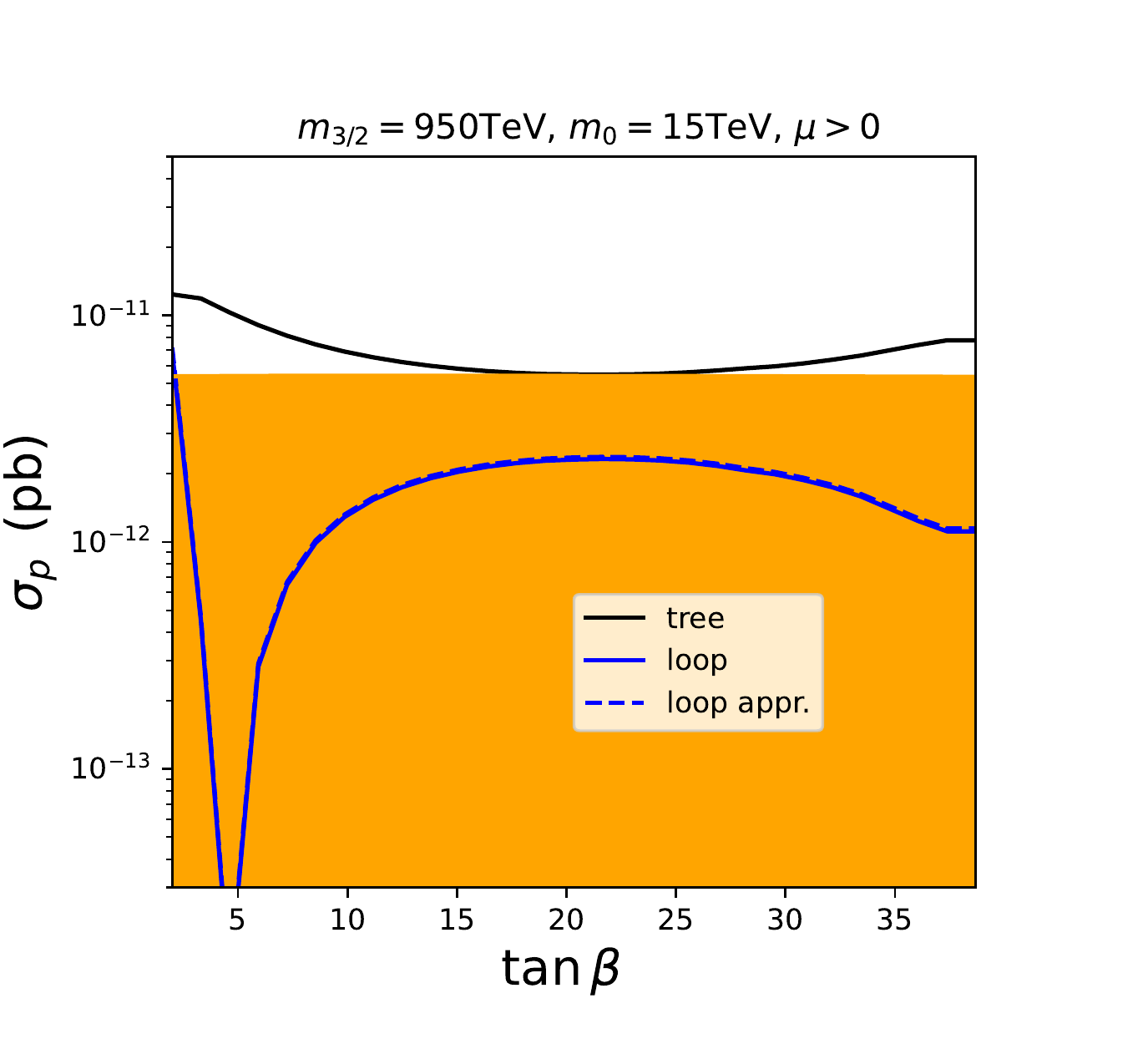}
\includegraphics[width=0.48\columnwidth]{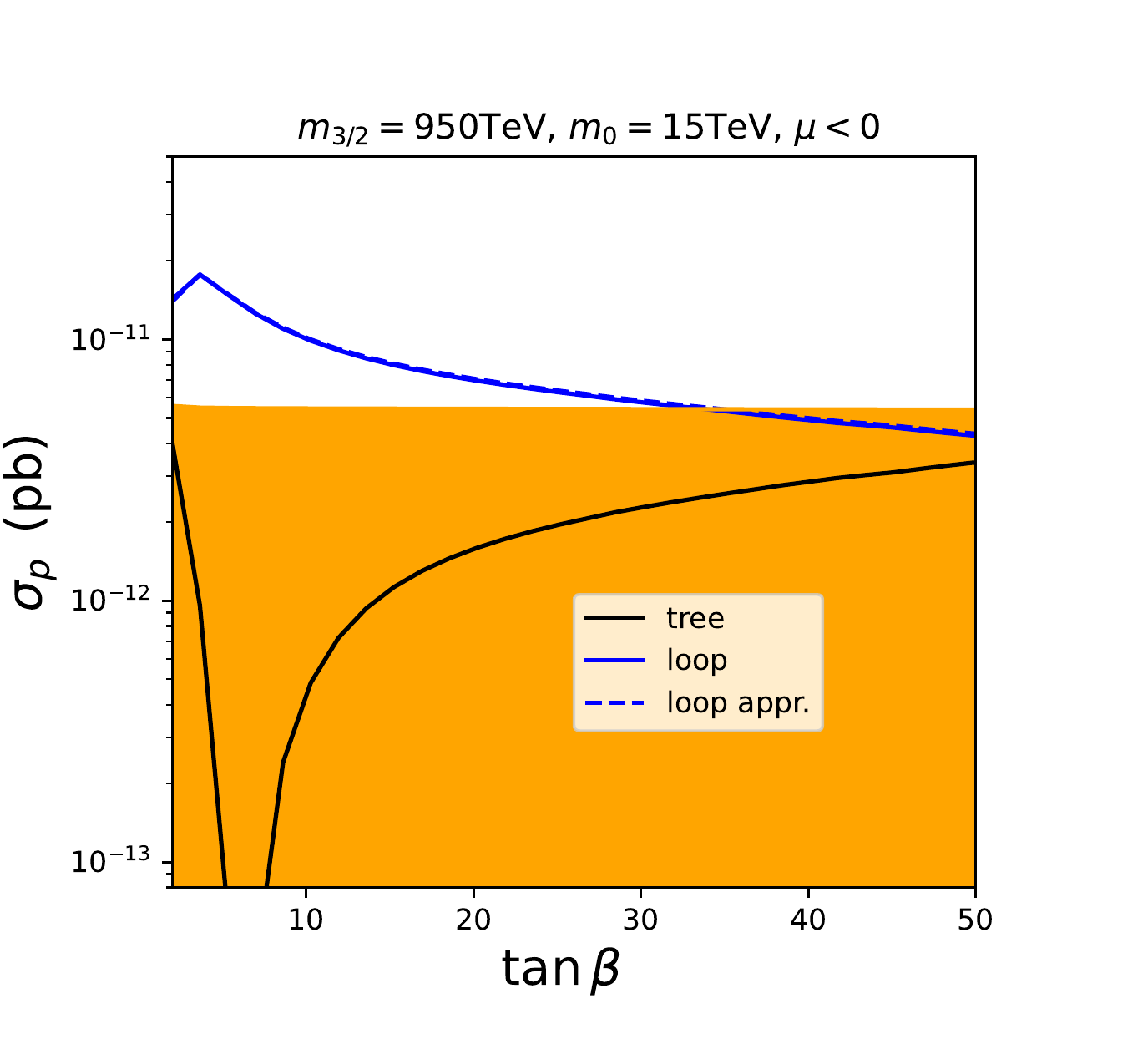} \\
\includegraphics[width=0.48\columnwidth]{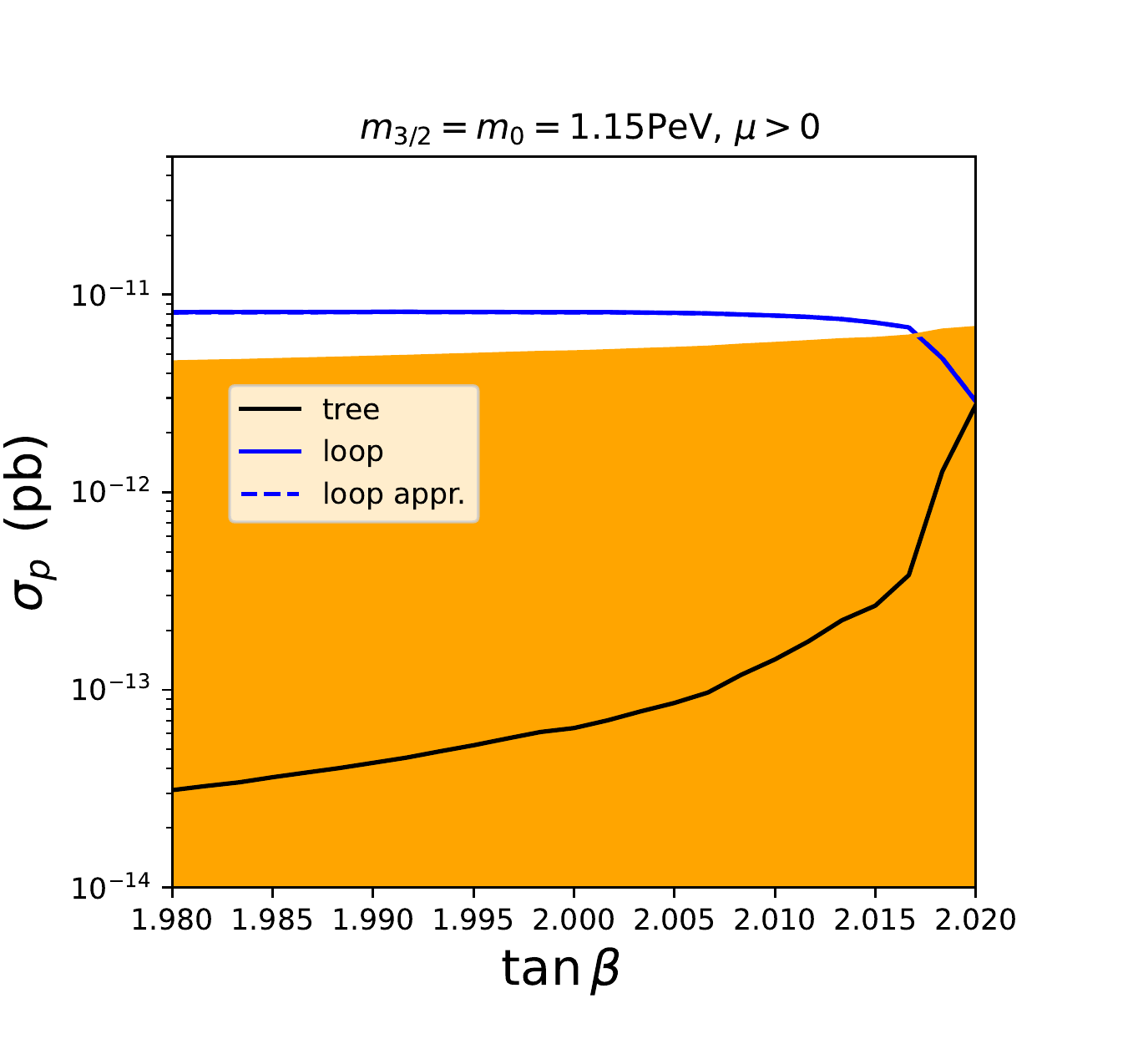}
\includegraphics[width=0.48\columnwidth]{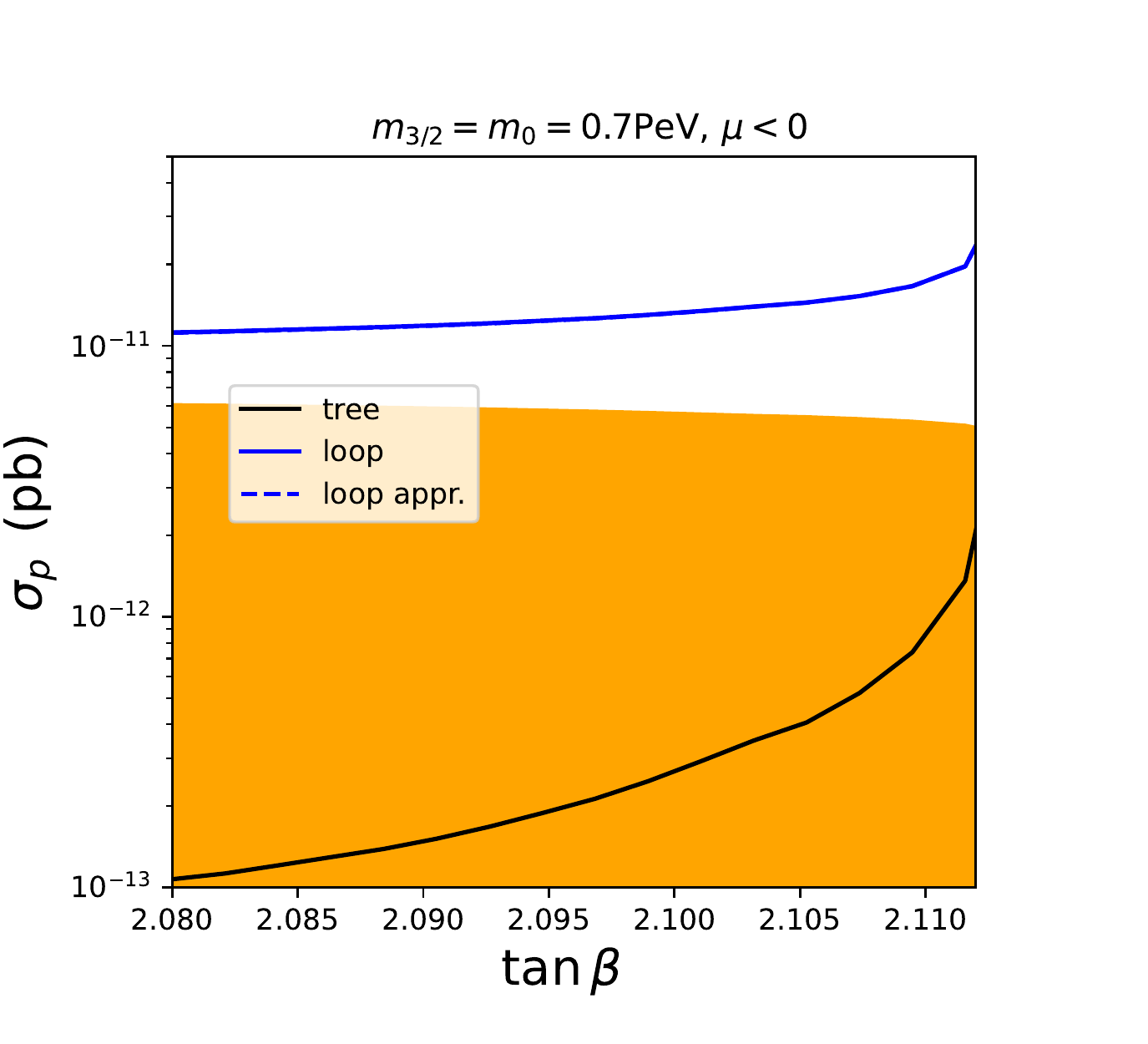}
  \caption{\it 
  Comparison of calculations of $\sigma_p$ as a function of $\tan \beta$
made using tree-level matrix elements (black lines) and
one-loop calculations made using the full expressions in the Appendix
(solid blue lines) and the simplified expressions in Eq.~(\ref{simplified})
that become exact when the wino neutralino mass $m_\chi \gg m_W$
(dashed blue lines).
The calculations are for the AMSB model with
$m_{3/2} = 950$ TeV, $m_0 = 15$ TeV and both signs of $\mu$ (upper panels) and the PGM-like models with $m_{3/2} = m_0 = 1.15$~PeV for $\mu > 0$ (lower left panel) and with $m_{3/2} = m_0 = 0.7$ PeV for $\mu < 0$ (lower right panel). The orange-shaded region corresponds to cross sections in the neutrino fog. In all four panels
the value of $m_{3/2}$ is chosen  to give $\Omega h^2 \simeq 0.12$.}
  \label{fig:tbsigmap}
\end{figure}

In the mAMSB model with $\tan \beta = 5$ and $\mu > 0$, we have the discouraging result that although the tree level result is above the neutrino fog (and hence in principle observable, though it is still two orders of magnitude below the current LZ limit), the one-loop correction induces a cancellation and the cross section drops into the neutrino fog unless $m_0 \gtrsim 23$ TeV.
Indeed, the cancellation is near-total when $m_0 = 16$ TeV. This cancellation can be understood by looking at the tree and loop contributions to the cross section separately. From Eqs.~\eqref{eq:a3aprx2} and \eqref{eq:fntree}, we see that $\alpha_{3q}/m_q$ is independent of $m_q$ for a wino LSP in mAMSB, and we can write
\begin{equation}
f_p^{(\rm tree)}\simeq 2\times10^{-6}{\rm GeV}^{-1}
\times
\frac{(M_2 + \mu \sin 2\beta)}{(M_2^2 -\mu^2)}
\, ,
\end{equation} 
when taking $m_h = 125$ GeV. When the relic density is fixed, requiring the wino mass, $M_2 \simeq 3$ TeV, and noting that $\mu$ (nearly equal to the Higgsino mass shown in Fig.~\ref{fig:mchimap}) varies with $m_0$, we see that the tree-level cross section varies little at low $m_0$ (as does $\mu$) and increases as $\mu$ begins to drop. This is the behaviour seen in the upper left panel of Fig.~\ref{fig:mchisigmap}.  When $M_2\simeq 3$\,TeV for wino DM and $m_h$ is restricted to the observed value, the loop correction is roughly constant and takes the value
\begin{equation}
	f_p^{(\rm loop)} \simeq 1.6\times 10^{-10}\,{\rm GeV}^{-2}\,.
\end{equation}
However, when $\mu > 0$ the tree level amplitude is negative and for a certain value of $\mu$ ($m_0$) will cancel the one-loop contribution as seen in Fig.~\ref{fig:mchisigmap}. It is easy to see that a complete cancellation, $f_p^{(\rm tree)}+f_p^{(\rm loop)}=0$, occurs when
\begin{equation}
	\sin 2\beta \simeq \frac{\mu}{12{\rm TeV}}-\frac{3.7{\rm TeV}}{\mu} \, .
\end{equation}
This occurs when $m_0 \sim 17$ TeV, when $\tan \beta = 5$ and $\mu \sim 9$ TeV as seen in Fig.~\ref{fig:mchimap}.
However, we note that the amplitude will not vanish if $|\mu|\gtrsim 15$~TeV.
This same cancellation can be seen in the upper panels of Fig.~\ref{fig:tbsigmap}, where we show the cross section as a function of $\tan \beta$ for $m_{3/2} = 950$ TeV and $m_0 = 15$ TeV.~\footnote{We note in passing that the results obtained using the
approximate expressions for the one-loop mass functions given in
Eq.~(\ref{simplified}) (dashed blue lines) are very similar to the
exact results, as was to be expected since $m_\chi \gg m_W$.
}

On the other hand, the tree-level scattering amplitude vanishes when
$M_2 + \mu \sin 2\beta=0$. 
For $\tan \beta = 5$, this occurs when $\mu \sim -8$ TeV, corresponding to $m_0 \sim 21$ TeV. 
However, when $\mu < 0$ the one-loop correction can enhances the total cross section, lifting it out of the neutrino fog and into the domain of possible experimental observation (though still unobservable at present). 
The large loop correction can significantly shift this point of vanishing amplitude. In the case of the upper right panel of Fig.~\ref{fig:mchisigmap}, 
the cancellation would occur at $m_0 > 25$~TeV, past the point where radiative electroweak symmetry breaking is possible.

In the PGM-like model with $\tan \beta = 2$ ($\sin 2\beta = 0.8$), $\mu \gg M_2$, and the tree-level amplitude is approximately proportional to $1/\mu$.
So long as $\mu$ is very large, the one-loop correction dominates the cross section substantially, lifting it out the neutrino fog unless $m_0 \gtrsim 1.2$ PeV when $\mu > 0$, as seen in the lower panels of Figs.~\ref{fig:mchisigmap} and \ref{fig:tbsigmap}. 

Before concluding, we note that the tree-level cross section in the mAMSB model was explored in a frequentist analysis using constraints from cosmology and accelerator experiments in~\cite{mc-amsb}. The best-fit cross section for wino dark matter was found to be slightly above (below) the neutrino fog for $\mu > 0$ ($\mu < 0$), and values for $\mu< 0$ extended far into the neutrino fog. 
It was also found that values of the cross section extend to much higher values. This occurs when $m_0$ takes large values and approaches the region with no electroweak symmetry breaking. This behaviour is seen, for example, in the upper left panel of Fig.~\ref{fig:mchisigmap} where the cross section increases by over an order of magnitude as $m_0$ is increased.

\section{Conclusion and Discussion}
\label{sec:conx}

The results in Figs.~\ref{fig:mchisigmap} and \ref{fig:tbsigmap} illustrate clearly the potential
importance of one-loop electroweak corrections to the spin-independent scattering cross section
for a wino-like LSP. In the case of the AMSB model, the results in the upper panels of these two
figures show sharp difference between the two signs of $\mu$. In the case of positive $\mu$
(upper left panels), the one-loop correction interferes negatively with the tree-level contribution
and has similar magnitude, potentially causing a total cancellation. On the other hand,
in the case of negative $\mu$ (upper right panels), the one-loop correction has the same sign as
the tree-level contribution for $m_0 \lesssim 20$~TeV and $\tan \beta \lesssim 5$, enhancing the spin-independent scattering cross section. Furthermore, in the case of the PGM-like model shown in the lower panels of Figs.~\ref{fig:mchisigmap} 
and \ref{fig:tbsigmap}, the one-loop corrections enhance the cross section for both signs of $\mu$.

The consequences of these effects is to invert the conclusions about experimental observability
that would be drawn from considering naively the tree-level cross section alone. In the case of 
the AMSB model shown in the upper panels of Figs.~\ref{fig:mchisigmap} and \ref{fig:tbsigmap}, 
whereas the tree-level calculation gives an observable cross section for $\mu > 0$
and a cross section lost in the neutrino fog for $\mu < 0$, the one-loop correction pushes
the cross section down into the fog for most values of $m_0$ and $\tan \beta$ when $\mu > 0$ and
lifts it out of the fog when $\mu < 0$. On the other hand, in the PGM-like model shown in the 
lower panels of Figs.~\ref{fig:mchisigmap} and \ref{fig:tbsigmap}, the effect of the one-loop
electroweak correction is generally positive, lifting the spin-independent cross section out of
the fog for most of the displayed ranges of $m_0$ and $\tan \beta$.

The one-loop electroweak corrections must therefore be taken into account when assessing the
implications of direct searches for the scattering of cold dark matter on nuclear targets for the
viability of AMSB and PGM-like models.~\footnote{As discussed in the Appendix, the full 
form of the one-loop correction is relatively complicated. However, the large mass of the wino
dark matter implies that it has a simple analytic approximation, see Eq.~(\ref{simplified}).} 
The good news is that the cross section is likely to be
above the neutrino fog for the AMSB model if $\mu < 0$ and for PGM-like models with either sign
of $\mu$. On the other hand, even in these encouraging cases the cross section is not much 
higher than the neutrino fog, so a strenuous experimental effort~\cite{XLZD} will be necessary to explore 
these models. However, the results of this paper indicate that this effort would have a good chance
of being rewarded if it is able to reach down to the level of the neutrino fog.

\section*{Acknowledgments}

The work of J.E. was supported by the United Kingdom STFC Grant ST/T000759/1.
The work of N.N. was supported by Grants-in-Aid for Scientific Research B (No. 20H01897) and Young Scientists (No.21K13916). 
K.A.O. was supported in part by DOE grant DE-SC0011842 at the University of Minnesota. J.Z. was supported in part by the NSF of China grants 11675086 and 11835005.

\section*{Appendix}
\appendix

We list here explicit expressions for the mass functions introduced
above:
\begin{align}
 g_{\rm H}(x)&= 2\sqrt{x}(2-x\ln x)-\frac{2}{b_x}(2+2x-x^2)\tan^{-1}\biggl(
\frac{2b_x}{\sqrt{x}}\biggr)~,\\
 g_{\rm B1}(x)&=\frac{1}{24}\sqrt{x}(2-x\ln x)+\frac{1}{24b_x}
(4-2x+x^2){\rm tan}^{-1}\biggl(\frac{2b_x}{\sqrt{x}}\biggr)
~,\\
g_{\rm B3}(x,y)&=
-\frac{x^{\frac{3}{2}} (2 y-x)}{12
   (y-x)^2}
-\frac{x^{\frac{3}{2}} y^3 \ln y }{24
   (y-x)^3}
+\frac{x^{\frac{5}{2}} (3 y^2-3
   x y+x^2 ) \ln x}{24
   (y-x)^3}
\nonumber\\
&+\frac{x^{\frac{3}{2}} \sqrt{y} (y^3-2
   y^2-14 y+6 x) \tan^{-1}\left(\frac{2 b_y}{\sqrt{y}}\right) }{24
   b_y (y-x)^3 }
\nonumber\\
&-\frac{x \left(x^4-3 y
   x^3-2 x^3+3 y^2 x^2+6 y
   x^2+4 x^2-6 y^2 x-6 y
   x-6 y^2\right) \tan
   ^{-1}\left(\frac{2 b_x}{\sqrt{x}}\right)}{24 b_x
   (y-x)^3} \ ,
\nonumber\\
 g_{\rm T_1}(x)&=\frac{1}{12}\sqrt{x}\{1-2x-x(2-x)\ln x\}
+\frac{1}{3}b_x(2+x^2)\tan^{-1}\biggl(\frac{2b_x}{\sqrt{x}}\biggr)
~,\\
g_{\rm T_2}(x)&=-\frac{1}{4}\sqrt{x}\{1-2x-x(2-x)\ln x\}+
\frac{1}{4b_x}x(2-4x+x^2)\tan^{-1}\biggl(\frac{2b_x}{\sqrt{x}}\biggr)~,\\
h_{{\rm T_1}}(x,y) &=
\frac{x^{\frac{3}{2}}\{x(1-2x)+y(13+2x)-2y^2\}}{12(x-y)^2} 
\nonumber \\ 
&-
\frac{x^{\frac{3}{2}}\{x^3(2-x)+2xy(3-3x+x^2)+6y^2(2-x)\}}{12(x-y)^3}
\ln x
\nonumber \\  &+
\frac{x^{\frac{3}{2}}y\{2x(3-6y+y^2)+y(12+2y-y^2)\}}{12(x-y)^3}
\ln y
\nonumber \\
&+ 
\frac{x\{4x^2b_x^2(2+x^2)-2xy(6-7x+5x^2-x^3)-6y^2(2-4x+x^2)\}}{12b_x(x-y)^3}
\tan^{-1}\left(\frac{2b_x}{\sqrt{x}}\right)
\nonumber \\ &-
\frac{x^{\frac{3}{2}}y^{\frac{1}{2}}\{2x(3-y)(2+5y-y^2)-y(2-y)(14+2y-y^2)\}}{12b_y(x-y)^3} 
\tan^{-1}\left(\frac{2b_y}{\sqrt{y}}\right)
,
\\
 h_{{\rm T_2}}(x,y) &=
 \frac{x^{\frac{3}{2}}\{x(-1+2x)-(1+2x)y+2y^2\}}{4(x-y)^2}
\nonumber \\ &+
\frac{x^{\frac{5}{2}}\{(2-x)x^2+2y(1-3x+x^2)\}}{4(x-y)^3}
\ln x
\nonumber \\ &+
\frac{x^{\frac{3}{2}}y\{y^2(y-2)-2x(1-3y+y^2)\}}{4(x-y)^3}
\ln y
 \nonumber \\ &+
\frac{x^3\{x(2-4x+x^2)-2y(5-5x+x^2)\}}{4b_x(x-y)^3}
\tan^{-1}\left(\frac{2b_x}{\sqrt{x}}\right)
\nonumber \\ &+
\frac{x^{\frac{3}{2}}y^{\frac{3}{2}}(2x(5-5y+y^2)-y(2-4y+y^2))}{4b_y(x-y)^3}
\tan^{-1}\left(\frac{2b_y}{\sqrt{y}}\right)~,
\end{align}
where we have defined $b_x\equiv \sqrt{1-x/4}$.

When the mass of the wino-like neutralino $m_{\chi}$ is much larger than the
electroweak scale; {i.e.,} $\omega, \tau \to 0$, the above mass
functions can be approximated as 
\begin{align}
 g_{\rm H}(\omega)&\simeq -2\pi \ ,
\nonumber\\
g_{\rm B1}(\omega)&\simeq \frac{\pi}{12} \ ,
\nonumber \\
g_{\rm B3}(\omega,\tau)&\simeq  \frac{\pi (2+3r)}{24(1+r)^3} ~,\nonumber\\
g_{\rm T1}(\omega)&\simeq \frac{\pi}3 \ ,
\nonumber\\
g_{\rm T2}(\omega)&\simeq  0 \ , 
\nonumber \\
 h_{\rm T1}(\omega,\tau) &\simeq \frac{\pi (2+3r)}{6(1+r)^3}~,
\nonumber\\
h_{\rm T2}(\omega, \tau)&\simeq  0 \ , 
\label{simplified}
\end{align}
with $r\equiv \sqrt{\tau/\omega}=m_t/m_W$.


\end{document}